\documentclass[referee,useAMS,usenatbib]{mn2e}
\usepackage{graphicx}
\title[INOV of RIQs]
  {Optical variability of radio-intermediate quasars}
\author[Goyal et al.]
  {Arti Goyal$^1$\thanks{E-mail: arti@aries.res.in},
   Gopal-Krishna$^{2}$,
   S. Joshi$^{1}$,
   R. Sagar$^{1}$,
   Paul J. Wiita$^{3,4}$,
\newauthor
   G. C. Anupama$^{5}$,
   D. K. Sahu$^{5}$,\\
$^1$ Aryabhatta Research Institute of Observational Sciences (ARIES),
Manora Peak, Naini Tal 263 129, India\\
$^2$ National Centre for Radio Astrophysics/TIFR, Pune University Campus, Pune 411 007, India\\
$^3$ School of Natural Sciences, Institute for Advanced Study, Princeton, NJ 08540, USA \\
$^4$ Department of Astronomy and Physics, Georgia State University, Atlanta, GA 30303, USA\\
$^5$ Indian Institute of Astrophysics (IIA) Bangalore 560 034, India \\
}
\date{Received 2009 XXXX XXX }

\pagerange{\pageref{firstpage}--\pageref{lastpage}} \pubyear{2007}

\def\LaTeX{L\kern-.36em\raise.3ex\hbox{a}\kern-.15em
    T\kern-.1667em\lower.7ex\hbox{E}\kern-.125emX}

\begin{document}

\label{firstpage}

\maketitle

\begin{abstract}
We report the results of our intensive intranight optical monitoring 
of 8 optically bright `radio-intermediate quasars' (RIQs) having  flat
or inverted radio spectra. The monitoring was carried out in 
{\it R-}band on 25 nights during 2005-09. On each night only one RIQ 
was monitored for a minimum duration of $\sim$ 4 hours (the average being 
5.2 hours per night).
Using the CCD as an N-star photometer, an intranight optical variability (INOV) detection threshold of 
$\sim$ 1--2\% was achieved for the densely sampled differential light 
curves (DLCs) derived from our data. These observations amount to a 
large increase over those reported hitherto for this rare and sparsely
studied class of quasars which can, however, play an important role in 
understanding the link between the dominant varieties of powerful AGN, 
namely the radio-quiet quasars (RQQs), radio-loud quasars (RLQs)  and 
blazars. Despite the probable presence of 
relativistically boosted nuclear jets, inferred from their flat/inverted 
radio spectra, clear evidence for INOV in our extensive 
observations was detected only on one night. 
Also, flux variation between two consecutive nights was clearly seen 
for one of the RIQs. These results demonstrate that as a class, RIQs are 
much less extreme in nuclear activity compared to blazars.
The availability in the literature of INOV data for another 2 RIQs 
conforming to our selection criteria allowed us to enlarge the
sample to 10 RIQs (monitored on a total of 42 nights for a minimum
duration of $\sim 4$ hours per night). The absence of large amplitude 
INOV $(\psi \geqslant 3\%)$ persists in this enlarged sample. This 
extensive database has enabled us to arrive at the first estimate 
for the INOV Duty Cycle (DC) of RIQs. The DC is
found to be small ($\sim$ 9\%), increasing to $\sim$ 14\% if the two 
cases of `probable' INOV are included. The corresponding value is known
to be $\sim 60\%$ for BL Lacs and $\approx 15\%$ for both RLQs
and RQQs, if they too are monitored for $\ga 4 - 6$ hours in each session. Our 
observations also provide information about the long-term optical variability (LTOV)
of RIQs, which is found to be fairly common and reaches typical 
amplitudes of $\approx$ 0.1-mag. The
light curves of these RIQs are briefly discussed in the context of a
theoretical framework proposed earlier for linking this rare kind 
of quasars to the much better studied dominant classes of quasars.

\end{abstract}


\section{Introduction}


%
%
%

\vskip0.5in

Although sparsely studied so far, radio-intermediate quasars 
(RIQs) can serve as an important tool for probing the relationship
between the radio-loud quasars (RLQs), blazars and radio-quiet 
quasars (RQQs, also termed as radio weak quasars). 
Ever since the discovery of quasars, the dichotomy in their radio 
loudness has been debated as an outstanding issue, since about
$\sim 10\%$ of optically selected quasars are found to be stronger
emitters in the radio band 
compared to the remaining population (Sandage 
1965; Strittmatter et al. 1980; Kellermann et al. 1989, 1994; 
Stocke et al. 1992).  Following
Kellermann et al. (1989) who made VLA observations of optically
selected Palomar-Green Bright Quasar Survey (BQS) objects, radio 
loudness is usually quantified in terms of a parameter `$R$', the 
ratio of continuum flux density at radio (5 GHz) and optical 
(4400 \AA) wavelengths. The histogram of $R$ found in their study
shows a large peak at $R \la 0.1-3$ and a second, weaker 
peak at $R \sim 100-1000$. Accordingly, quasars falling in the
range $3 < R < 100$ can be termed as ``Radio Intermediate Quasars 
(RIQs)'' (e.g., Miller, Rawlings \& Saunders 1993; see also, 
Diamond-Stanic et al. 2009). Although it has been argued that 
RLQs are merely the tail of a broad distribution of quasar radio 
strengths and not a separate population (e.g., Cirasuolo et al. 2003) 
careful analysis of the largest available samples strongly indicates 
that there is indeed a bimodal distribution in $R$ 
(Ivezi\'c et al. 2002, 2004) and that the radio loud fraction increases 
with rising optical luminosity but decreases with increasing redshift 
(e.g., Jiang et al. 2007; Rafter, Crenshaw \& Wiita 2009).

The observed radio luminosity of a RIQ is typically well above the 
value dividing the Fanaroff-Riley type I (FRI) and type II (FRII) 
radio galaxies [log$P_{20cm}$ (W/Hz) $\simeq 25.0$] (Falcke, 
Gopal-Krishna \& Biermann 1995b), but probably not so once the 
possibility of relativistic flux boosting is taken into account (e.g., Wang et al. 2006).
It has been long been suggested that RIQs are Doppler boosted counterparts of RQQs, 
such that the relativistic jet is closely aligned to the line of
sight. This was mainly inferred from the radio versus [O III] line
intensity diagram for optically selected quasars and also from the
high brightness temperature of the radio emission from RIQs, as well as their 
usually flat or inverted radio spectra and substantial radio flux variability 
(Miller, Rawlings \& Saunders 1993; Falcke, Gopal-Krishna \& Biermann 1995b;
Falcke, Patnaik \& Sherwood 1996a; 
Xu et al. 1999; cf. Barvainis et al. 2005; Falcke et al. 2001). 
 
Implicit in this paradigm is the presence of relativistic jets in 
RQQs. This premise has received considerable support from deep radio 
imaging, which has revealed faint kpc-scale radio structures 
in several RQQs (e.g., Kellermann et al. 1994; Blundell \& Rawlings 2001; Leipski et al. 2006). 
In addition, Blundell, Beasly \& Bicknell 
(2003) have reported evidence for a highly relativistic parsec scale radio 
jet in the RQQ PG1407+263. Independent support for the possibility of RQQs 
possessing relativistic jets has come from the observations of intranight 
optical micro-variability in RQQs (e.g., Gopal-Krishna et al. 2000, 2003; 
Gupta and Joshi 2005; Czerny et al. 2008) and in their weaker cousins, 
Seyfert 1 galaxies (Jang \& Miller 1995, 1997; Carini, Noble \& Miller 2003). At least a modest amount 
of relativistic beaming in the jets of radio weak quasars has also 
been inferred from the detection of compact cores on VLBI scales and 
radio variability on 
month-like to year-like timescales, indicating minimum brightness temperatures 
in a broad range from $10^8$--$10^{11}$ K  (e.g., Ulvestad, Antonucci \& Barvainis 2005; 
Barvainis et al. 2005; Wang et al. 2006).  
However, there is at present little clarity about the nature of jets 
in RQQs {\it vis-\'a-vis} those in RLQs and blazars whose jets are
widely attributed to spinning super massive black holes (SMBHs) (Blandford 
1990 and references therein; Wilson \& Colbert 1995), or more
massive SMBHs accreting at smaller fraction of the Eddington rate
(e.g., Boroson 2002). 
A number of authors have drawn an analogy between the radio 
emission of quasars and X-ray binaries and have attributed the main 
difference between RLQs and RQQs to different accretion modes
changing the nature of their jets (e.g., 
K\"{o}rding, Jester \& Fender 2006). Stellar mass black holes are 
known to slide into a state where radio emission is strongly 
suppressed or quenched (e.g., Corbel et al. 2001; Maccarone, Gallo \& 
Fender 2003). 
In this general picture RIQs 
could well be relativistically beamed counterparts of the RQQ jets. 
In alternative schemes, the radio weakness of RQQs is attributed to a 
rapid deceleration of the jet on subparsec scale, caused perhaps by: 
strong interaction with the torus material (e.g., Falcke, Malkan \& Biermann 1995a; 
Falcke, Gopal-Krishna \& Biermann 1995b); mass loading with
the debris of stars tidally disrupted by the SMBH (Gopal-Krishna, Mangalam
\& Wiita 2008); ejection velocities less than the escape velocity 
(Ghisellini, Haardt \& Matt 2004); or a high photon density 
environment produced by the quasar  (Barvainis et al. 2005). 

If indeed, RIQs are relativistically boosted counterparts of RQQs,
observing the former provides an opportunity to better detect RQQ jets and 
compare their properties with the better studied, more powerful radio 
jets of RLQs. For instance, it would be interesting to enquire if
the postulated relativistic beaming of RQQ jets manifests itself as
blazar-like behaviour in RIQs. Indeed, this seems to be the case for
the most prominent example of the RIQ class, namely III Zw 2 (see below).
However, essentially no such information is currently available for 
RIQs as a class even though they, despite being numerically rare, are 
a potentially very important link between the major AGN classes powered
by nuclear activity, namely RQQs, RLQs and blazars. The present study 
is a first attempt to bridge this gap in a statistically significant 
manner.

In terms of radio properties, RIQs form a distinct subclass. Their
dominant feature is a radio core (compact on arcsecond scale) of a
flat, variable radio spectrum (Falcke, Patnaik \& Sherwood 1996a; Kukula et al.
1998; also, Kellermann et al. 1994; Ulvestad et al. 2005). 
The aforementioned view that they are relativistically boosted RQQ 
jets has found strong support from the discoveries of long-term, large
amplitude ({\bf up to a factor of 20}) radio variability of the RIQ III Zw 2 
(B0007$+$106; Ter\"asranta et al. 1998) and of superluminal motion 
in its radio core (Brunthaler et al.\ 2000, 2005), both being canonical 
attributes of blazars. 
Unfortunately, equivalent observational data are severely lacking 
for other known RIQs. It may further be emphasized that III Zw 2 
($z = 0.089$) has in fact been classified as Seyfert 1 (Arp 1968; 
Osterbrock 1977) hosted by a spiral galaxy (Hutchings \& Campbell 1983; 
Taylor et al. 1996). Thus, an interesting question is which, if any, 
blazar-like characteristics are associated with RIQs of high optical 
luminosity ($ M_{B} < -23.5$), which are bona-fide QSOs and hence more 
likely to be associated with massive elliptical galaxies. 
Promising indications come from the recent finding that for the radio
spectra of the RIQs found in the SDSS are typically flat or inverted
(consistent with the postulated relativistic beaming of their jets) and, 
moreover, radio flux variability (near 1 GHz) on year-like time scales 
is actually comparably common for RIQs and RLQs 
(Wang et al. 2006). For both these reasons, RIQs appear to be promising candidates for INOV, 
albeit it is unclear if the mechanism responsible for long-term 
variations is also responsible for INOV.

In recent years, INOV has been increasingly recognized as a signature 
of blazar-like jets.  Extensive observations have shown that INOV with
large amplitude (with variations, $\psi$, exceeding 3\%) occurs almost 
exclusively in blazars and, furthermore, the duty cycle (DC) of INOV 
in blazars is very high ($> 50\%$), provided the monitoring duration 
exceeds about 4 hours (e.g., Carini 1990; Gopal-Krishna et al. 2003; Stalin et
al. 2004a, 2005). To exploit this clue we 
have carried out an INOV survey of a fairly large and representative 
sample of 8 RIQs, further augmented by the INOV data available in
the literature for another 2 RIQs that meet the selection criteria
adopted for our sample of 8 RIQs. Thus, the enlarged sample consists 
of 10 optically bright and intrinsically luminous, core-dominated
RIQs. 
{\bf Cosnsiderable evidence exisits for relativistic beaming in these
RIQs. Falcke, Mathew \& Biermann (1995a) find J1336+1725 to be radio 
variable, but not J1701+5149. For J1259+3423 multi-epoch flux measurements 
are not known to us, hence its variablity status is
unknown at present. The remaining 7 RIQs in our sample are all found
to be radio variable (Wang et al. 2006, Falcke, Sharwood \& Patnaik 1996). 
Thus, at least 8 of the 10
RIQs are known to shown radio variability. Furthermore, spectral
information is available in the literature for 8 out of the 10 RIQs
and each case the radio spectrum is found to be either flat or inverted (see Table 1). 
All this suggests that the radio jets in essentially
all there RIQs are at least modestly Doppler boosted.
}.

\section{The RIQ sample}

Since detection of any blazar-like jet characteristics in RIQs is our 
main objective here, we have focused on the RIQs for which  credible 
evidence for relativistically boosted jet emission exists 
(at least in the radio band). 
This leads us to largely excluding the steep-spectrum RIQs, which also exist 
(e.g., Mart{\'i}nez-Sansigre et al.\ 2006). Thus, our sample contains only RIQs 
having flat or inverted radio spectra (at least a flat-spectrum radio core), 
in addition to satisfying the usual criterion, namely, {\it K}-corrected ratio $ {\it R}^{*}$
of 5 GHz to 2500 \AA \ fluxes, falling in the range
3 $<$ {\it R}$^{*}$ $<$100 (Section 1; Miller, Rawlings \& Saunders 1993). 
Moreover, in 
order to attain an INOV detection threshold of $\sim$1--2\% using a 1--2 
metre class telescope equipped with a CCD detector, we have confined 
ourselves to optically bright RIQs ($m_B$ $<$ 18.0 mag). Finally, 
in order to minimize contamination from the host galaxy (Cellone, 
Romero \& Combi 2000), our sample was restricted to intrinsically 
luminous AGNs with $M_{B}$ $<$ $-$23.5 mag (see also Stalin et al. 2004a). 
Based on these well defined basic criteria, our sample (Table 1) was
assembled in the following manner:

(a) Seven RIQs with adequate brightnesses and good nearby 
comparison stars originally
were selected from the list of 89 RIQs prepared by 
Wang et al. (2006) based on the detection of radio flux variability 
among SDSS quasars; they are thus presumed to have a flat/inverted 
radio spectrum. Out of these, three RIQs (J140730.43$+$545601.6, 
J162548.79$+$264658.7 and J210757.67$-$062010.6) could not be monitored
due to observing time constraints, leaving the remaining 4 RIQs in
our sample.
 
(b) Four RIQs were chosen from  Table 1 of Falcke, Sherwood \& 
Patnaik (1996b) on the criterion of having a flat/inverted radio 
spectrum, based on their quasi-simultaneous measurements at 2.7 and 
10 GHz ($\alpha$ $> -$0.5; $S_{\nu}$ $\propto$ $\nu^{\alpha}$).
 
(c) One RIQ (J1259+3423) was contributed by the sample observed by Carini et al.
(2007).
  
(d) One RIQ (J1701$+$5149) was included in our sample, despite its 
being reported a as steep-spectrum type (Falcke, Sherwood \& Patnaik 1996b),
because it was later  resolved and found to possess a 
core with a flat spectrum, with $\alpha_{4.8}^{8.4}$ 
$\simeq -0.2$ (Kukula 
et al.\ 1998; see also, Hutchings, Neff \& Gower 1992). 
 
Lastly, we note that the RIQ J0832$+$3707 is included in the our sample 
inspite of its being nearly 1-mag fainter than the adopted threshold 
($M_{B}$ $<$ $-$23.5 mag), since it is found to be stellar in the FBQS 
survey (White et al.\ 2000). 
However, its exclusion from the sample would have no significant 
impact on our conclusions.
Table 1 lists our sample of 10 RIQs. Columns are as follows: 
(1) source name (an asterisk indicates that the source was monitored
by us); (2) right ascension; (3) declination; (4) apparent B-magnitude; 
(5) absolute B-magnitude; (6) redshift; (7) optical polarization; (8)
 radio luminosity at 5 GHz; (9) radio spectral index; (10) radio loudness 
parameter R computed by us; (11) reference code. We have used a concordance 
cosmology with Hubble constant $H_{0} = 71$ km sec$^{-1}$ Mpc$^{-1}$,
$\Omega_{M} =0.27$ and $\Omega_{\lambda} =0.73$ (Komatsu et al.\ 2009)
and compute the luminosity distance using Hogg (1999).

\section{Observations}
\subsection{Instruments used}
The observations were mainly carried out using the 104-cm
Sampurnanand telescope (ST) located at
Aryabhatta Research Institute of observational sciencES (ARIES),
Naini Tal, India. It has a Ritchey-Chretien (RC) optics with
a f$/$13 beam (Sagar 1999). The detector was a cryogenically
cooled 2048 $\times$ 2048 chip mounted at the Cassegrain focus.
This chip has a readout noise of 5.3 e$^{-}$/pixel and a gain 
of 10 e$^{-}$$/$Analog to Digital Unit (ADU) in the usually 
employed slow readout mode. Each pixel has a dimension of 
24 $\mu$m$^{2}$, corresponding to 0.37 arcsec$^{2}$ on 
the sky, thereby covering a total field of 13$^{\prime}$ $\times$
13$^{\prime}$. We carried out observations in a 2 $\times$ 
2 binned mode to improve the S$/$N ratio. 
The seeing usually ranged between
$\sim$1$^{\prime\prime}$.5 and $\sim$3$^{\prime\prime}$.0, 
as determined using 3 fairly bright stars on the CCD frame; 
plots of the seeing are provided for all of the nights
in the bottom panels of Figure 1 (see Section 4).

Some of the observations were carried out using 200-cm IUCAA Girawali
Observatory (IGO) telescope located at Girawali, Pune, India, which is a RC design
with a f$/$10 beam at the Cassegrain focus\footnote
{http://www.iucaa.ernet.in/\%7Eitp/igoweb/igo$_{-}$tele$_{-}$and$_{-}$inst.htm}. 
The detector was a cryogenicallly cooled 2110$\times$2048 chip mounted 
at the Cassegrain focus. The pixel size is 15 $\mu$m$^{2}$ so that the 
image scale of 0.27 arcsec$/$pixel covers an area on 10$^{\prime}$ 
$\times$ 10${^\prime}$ on the sky. The readout noise
of CCD is 4.0 e$^{-}$/pixel and the gain is 1.5 e$^{-}$$/$ADU. 
The CCD was used in an unbinned mode. The seeing ranged mostly 
between $\sim$1$^{\prime\prime}$.0 and $\sim$3$^{\prime\prime}$.0.

The other telescope used by us for monitoring the RIQs is the 201-cm 
Himalayan Chandra Telescope (HCT) located at Indian Astronomical 
Observatory (IAO), Hanle, India. It is also of the RC design 
with a f$/$9 beam at the
Cassegrain focus\footnote{http://www.iiap.res.in/$\sim$iao}.
The detector was a cryogenically cooled 2048 $\times$ 4096 chip, 
of which the central 2048 $\times$ 2048 pixels were used. 
The pixel size is 15 $\mu$m$^{2}$ so that the image scale of 
0.29 arcsec$/$pixel covers an area of about 10$^{\prime}$ $\times$ 10${^\prime}$ 
on the sky. The readout noise of CCD is 4.87 e$^{-}$/pixel and the 
gain is 1.22 e$^{-}$$/$ADU. The CCD was used in an unbinned mode. 
The seeing ranged mostly between $\sim$1$^{\prime\prime}$.0 to 
$\sim$4$^{\prime\prime}$.0.\\ 

All the observations were made using {\it R} filter as in 
this band these CCDs have maximum response. 
The exposure time was typically 12--30 minutes for the 
ARIES and IGO observations and ranged from 3 to 6 minutes for 
observations from IAO, depending on the brightness of the source, 
phase of moon and the sky transparency for that night. 
The field positioning was adjusted so as to also have within the
CCD frame 2-3 comparison stars within about a magnitude of the RQQ, 
in order to minimize the possibility of getting spurious 
variability detection (e.g., Cellone, Romero \& Araudo 2007). 
For all three telescopes bias frames were taken
intermittently and twilight sky flats were obtained.

\subsection{Data reduction}
The preprocessing of images (bias subtraction, flat-fielding 
and cosmic-ray removal) was done by applying the regular 
procedures in IRAF\footnote{\textsc {Image Reduction
and Analysis Facility}} and MIDAS\footnote{\textsc 
{Munich Image and Data Analysis System}} software. 
The instrumental magnitudes of the RIQ and the stars 
in the image frames were determined by aperture photometry, 
using {\textsc {DAOPHOT}} \textrm{II}\footnote{\textsc {Dominion
Astrophysical Observatory Photometry} software} (Stetson 1987).
The magnitude of the RIQ was measured relative to the nearly 
steady comparison stars present on the same CCD frame 
(Table 3). This way, differential light curves (DLCs) of 
each RIQ were derived relative to 2 or 3 comparison stars.
For each night, the selection of optimum aperture radius 
was done on the basis of the observed dispersions in the 
star-star DLCs for different aperture radii starting from 
the median seeing (FWHM) value on that night up to 4 times 
that value.  The aperture selected was the one which 
showed minimum scatter for the steadiest DLC found for 
the various pairs of the comparison stars (e.g., Stalin et al. 2004a). 

\section{Results}
\subsection{Differential Light Curves}
Figure 1 shows the intranight DLCs obtained for RIQs monitored in the 
present study. We later combined each intranight DLC for a particular RIQ 
and produced its long term optical variability (LTOV) DLC relative to 
the same set of steady stars that we had used for making the intranight DLCs. 
These LTOV DLCs are shown in Fig. 2.
In Table 2, we summarize the observations 
for the entire sample of RIQs including the two RIQs
added from the literature. For each night of observation
we list the object name, date of monitoring, telescope 
used, duration of the observation, number of data points 
($N_{points}$) in the DLC, average rms of the pairs of star-star DLC, the INOV 
amplitude ($\psi$) and C$_{eff}$, an indicator of 
variability status and the reference for the INOV data.
The classification `variable' (V) or `non-variable' 
(N) was decided using the parameter C$_{eff}$, 
basically defined following the criteria of
Jang \& Miller (1997). We define C for a given DLC 
as the ratio of its standard deviation, $\sigma$$_T$, 
and $\eta\sigma_{err}$, where $\sigma_{err}$ is the 
average of the rms errors of its individual 
data points and $\eta$ was estimated to be 1.5 
(Stalin et al. 2004a,b, 2005; Gopal-Krishna 
et al. 2003; Sagar et al. 2004). However, our analysis 
for the present dataset yields $\eta = 1.3$ and we
have adopted this value here. We compute C$_{eff}$
from the C values (as defined above) found for the 
DLCs of an AGN relative to different comparison stars 
monitored on a given night (details are given in
Sagar et al. 2004). This has the advantage of using 
multiple DLCs of an AGN, relative to the different 
comparison stars. The source is termed `V' for 
C$_{eff}$$ > $ 2.576, corresponding to a confidence
level of $>$99\%. We call the source a `probable 
variable' (PV) if C$_{eff}$ is in range of 
1.950 to 2.576, corresponding to a confidence
level between 95\% to 99\%. Finally, the peak-to-peak 
INOV amplitude is calculated using the definition
(Romero, Cellone \& Combi 1999) 

\begin{equation}
\psi= \sqrt{({D_{max}}-{D_{min}})^2-2\sigma^2}
\end{equation}
with \\
$D_{max}$ = maximum in the AGN's DLC  \\
$D_{min}$ = minimum in the AGN's DLC \\
$\sigma^2$= $\eta^2$$\langle\sigma^2_{err}\rangle$.

\subsection{The INOV duty cycle (DC)}

The INOV DC for our entire sample of RIQs (Table 2) 
was computed following the definition of Romero, Cellone \& Combi (1999)
(see, also, Stalin et al. 2004a):

\begin{equation}
DC  = 100\frac{\sum_{i=1}^n N_i(1/\Delta t_i)}{\sum_{i=1}^n (1/\Delta t_i)}\%
\label{eqno1}
\end{equation}
where $\Delta t_i = \Delta t_{i,obs}(1+z)^{-1}$ is the
duration of monitoring session of a RIQ on the $i$th night, corrected for
the RIQ's cosmological redshift, $z$; $N_i$ was set equal to 1 if
INOV was detected, otherwise $N_i$ = 0. Note that since the duration of
monitoring of a given source was not the same on all the nights, 
the computation has been weighted
by the actual duration of monitoring, $\Delta t_i$. 

Although the data taken from the
literature was in some cases in V band instead of R band, for the present
purpose we do not distinguish between the V and R bands.
In this manner, we computed the INOV DC for our entire data set of 42 nights. 
It was found to be only $\sim 9\%$, which increases to $\sim 14\% $ if the
two cases of probable INOV are also included (Table 2). Note that even for `P' and
``PV'' cases, the INOV amplitude 
always remained modest, with $\psi <$ 3\%. 

As noted in Table 1, optical polarization measurements are available for 
6 of our total 10 RIQs and in each case $P_{op} < 1\%$, while the 
value that nominally defines the highly polarized AGNs 
is $P_{op} > 3\%$ (e.g., Impey \& Tapia 1990); therefore at least these 6 RIQs 
lack a strong blazar-like synchrotron component in the optical band despite having
flat/inverted radio spectrum (Table 1).
It is worth recalling that for single epoch measurements
about 90\% of radio selected BL Lacs show $P_{op} > 3\%$, while about
half of the X-ray selected BL Lacs evince that same high polarization 
level (Jannuzi, Smith \& Elston 1994).
Therefore only rarely will an intrinsically highly polarized object be 
observed as a lowly polarized object.

\section{Notes on individual sources}

{\bf RIQ J0748+2200:} We monitored this RIQ on four nights spanning 
about a year. Formally significant variations with $\psi \sim$ 2.3\% 
and 1.4\%, were seen on 23 Jan. 2007 and 30 Jan. 2008, respectively. 
Note that this RIQ has a contaminating nearby faint object 
$\sim$ 8$^{\prime\prime}.0$ offset along PA $\sim$ 229$^\circ$. 
Still we have carried out aperture photometry, following the argument
(Howell 1990) that for moderately clustered objects, i.e., those separated 
by $\geq$2$\times$FWHM from the companion, the technique of optimum 
extraction based on aperture growth curve method can be taken as a viable 
alternative to the traditional crowded point source photometry. 

The mild variability noticed on 23 Jan. 2007 might be an artefact of
rather poor seeing (which varied between to $2^{\prime\prime}.5$ 
and $3^{\prime\prime}.5$). Indeed the seeing variation (plotted 
at the bottom of the DLCs in Fig. 1) does seem partially correlated 
with the observed  variation of the RIQ-star DLCs. Therefore, 
we prefer to designate this RIQ as probably-variable (PV) on this 
night (despite $C_{eff} =$ 2.58). A high statistical
significance was also estimated for the INOV seen on  30 Jan. 2007,
when the RIQ brightened by $\sim$ 1.4\% (Fig. 1). Although in this case 
too, the brightening coincides with the time when the seeing changed from 
$2^{\prime\prime}.0$ to $2^{\prime\prime}.5$ (between UT 18.0 to 19.5 
hours), we do not see any general correlation between the variations of
seeing and the source magnitude and there is also a very high value
of $C_{eff} = 5.1$ for this night. Taking this into 
account we have designated this RIQ as
variable (V) on this night (Table 2). 
But on the longer-term, this RIQ showed no variability,
with $\psi <$ 0.02 mag over the time span of $\sim$ 1 year (Fig. 2).
Note, however, that the comparison star S1 showed a brightness dip by
$\sim$ 0.03 mag over that period.\\

{\bf RIQ J0832+3707:} We monitored this RIQ on four nights spanning seven
weeks. No INOV was detected; however, this RIQ remains thus far our best 
case of  internight variability, 
showing a fading by $\sim$ 0.06 mag between 21 Feb.\ 2007 and 10 Mar.\ 2007
and a brightening of $\sim 0.02$ mag by the following night (Fig. 2). \\

{\bf RIQ J0836+4426:} This RIQ did not show INOV on any of the 
three nights it was monitored over the time span of 45 days.  
In the longer-term, it showed a 
fading by 0.05 mag, between 22 Jan. 2007 and 10 Feb. 2007, followed by 
a $\sim$0.02 mag brightening at the time of its last observations on 
9 Mar.\ 2007. Here too, the S1 varied by about 0.04 mag between the first
and second dates it was observed.\\

{\bf  RIQ J0907+5515:} We monitored this RIQ on two consecutive nights, 
but found no INOV down to a limit of 0.02 mag. Also, no variability was 
detected between the two nights. \\ 

{\bf RIQ J1259+3423:} Out of the three nights we monitored this RIQ, it
is classified as ``probable variable'' on the night of 20 Apr. 2007.
Earlier, Carini et al. (2007) monitored it for four nights during 
1998. While they did not detect significant intranight fluctuations, 
internight variability of $\sim 0.2$ magnitude was detected by 
them over the course of 60 hours. In our observations spanning six days, 
this RIQ showed no internight variability. \\

{\bf RIQ J1336+1725:} This RIQ remained non-variable on all the three nights
it was monitored by us in the course of three years. In the longer term, it 
has shown a moderate fading by $\sim$ 0.03 mag over a year
(Fig. 2). \\ 

{\bf RIQ J1539+4735:} Based on the three nights' monitoring over the time 
span of 18 days, no significant INOV or LTOV was detected (Figures 1 \& 2). It had 
earlier been monitored by Jang (2005) for a single night but, again, no
INOV was detected (Table 2). \\

{\bf RIQ J1719+4804:} We monitored this RIQ on three nights over a month,
but no INOV was detected. Likewise, internight change was also not
found between  29 Apr.\ 2006 and the following night. However, a brightening 
by 0.04 mag was observed between 30 Apr. 2006 and 30 May 2006 (Fig. 2). 
Jang \& Miller (1995) monitored this RIQ on two nights during 1994,
for durations of 3.3 and 3.8 hours. They report confirmed INOV on the
former night (Table 2). Even though their monitoring durations fall marginally 
short of our criterion, we have included their data in
the present study (Table 2).\\

For the two RIQs not covered in our monitoring programme, the summary of 
INOV and LTOV results, taken from the literature, is as follows:\\ 

{\bf RIQ J1312+3515:} This RIQ was monitored by Carini et al. (2007) 
for three nights but no INOV was detected (Table 2). They do not comment on
any longer-term fluctuations over the four-days' time span covered by their
observations. Although INOV remained undetected also in the three nights' 
monitoring of this RIQ by Sagar et al. (2004) (Table 2), a 0.10 mag
fading was found between the first two epochs of their monitoring programme, 
separated by two years.

{\bf RIQ J1701+5149:} This RIQ was monitored on four nights by Carini et al.
(2007), but no INOV was seen (Table 2). Also, significant fluctuations 
were found to be absent both on inter-night or longer timescales.
This RIQ was also monitored by Jang (2005) who detected a confirmed 
INOV ($C_{eff}= 3.1$, Table 2).

\section{Discussion and Conclusions}

The results reported here for a well defined, representative sample of 
{\bf flat/inverted spectrum} RIQs provide a large increase over the existing information, both in 
terms of sample size and observing time. Thus, they permit the first 
good estimate of the INOV characteristics of RIQs and allow their 
comparison with those of the major AGN classes that are widely
separated in the degree of radio loudness. The INOV mechanism for the major 
AGN classes continues to be debated (Section 1). For the (jet-dominated) 
blazars, INOV is believed to be associated with irregularities in the 
non-thermal Doppler boosted jet flow, impacted by shocks (e.g., 
Blandford \& K{\"o}nigl 1979; Miller, Carini \& Goodrich 1989; 
Marscher 1996). In contrast, 
the instabilities or perturbations within the accretion disc might 
contribute very significantly to the INOV in the case of RQQs 
(e.g., Wiita et al. 1991; Mangalam \& Wiita 1993),
particularly since any contribution from the jet must be weak. 

To put the question of INOV of RIQs in perspective one 
may recall the known INOV characteristics of blazars (e.g., Heidt \& Wagner 1996; 
Dai et al. 2001; Romero et al. 2002; Xie et al. 2002; Sagar et al. 2004; 
Stalin et al. 2005) and RQQs (e.g., Gopal-Krishna et al. 1993, 2003; 
Jang \& Miller 1995, 1997;  Stalin et al. 2005; Carini et al. 2007). 
The major findings of these studies are that blazars tend to vary 
more frequently and more strongly ($\psi > 3\%$) on intranight timescales,
in stark contrast to RQQs and even to non-blazar RLQs, which show only mild 
INOV (i.e., $\psi < 3\%$; Stalin et al. 2004b). Moreover, for monitoring 
durations in excess of $\sim 4$ hours, the INOV duty cycle is $\ga 60\%$ 
for BL Lacs (Carini 1990; Miller \& Noble 1996; Romero et al. 2002; 
Gopal-Krishna et al. 2003; Stalin et al. 2005), mostly with large INOV
amplitudes ($\psi > 3\%$; Gopal-Krishna et al. 2003) but the duty
cycle is only $\la 15\%$ for RQQs (Jang \& Miller 1995, 1997; de Diego et 
al. 1998;
Romero et al. 1999; Gopal-Krishna et al. 2003; Stalin et al. 2004b),
or even core dominated RLQs (Stalin et al. 2004b; Ram{\'i}rez et al. 2009).

Thus, the key result from the present study is that the INOV duty cycle 
for RIQs is small ($\sim$ 10\%), and clearly not greater than that known for 
RQQs and non-blazar type RLQs.  Although exact values of computed DCs will
depend upon the sample and analysis technique, the key difference in
INOV frequency is between blazars and  both radio loud and
radio quiet non-blazar classes (Stalin et al. 2004b, 2005) as 
also confirmed recently by Ram{\'i}rez et al. (2009). Moreover, on no occasion (among the 42
nights) do we have the evidence for an INOV amplitude exceeding $3\%$ 
level, even though each of the 10 RIQs was monitored for at least 
$\sim$ 4 hours in every session (Sect.\ 1; Table 2).
Specifically, the monitoring durations in our 
programme ranged between 3.9 hours to 7.6 hours, with 
an average of 5.2 hours. As emphasized by Carini (1990), the  duration of 
monitoring in a given session can play a substantial role in determining the 
INOV status of an AGN, such that the probability of confirmed INOV 
detection in BL Lacs is found to increase from $50\%$ to $80\%$ if the 
duration of monitoring is raised from $\sim$ 3 hours to $\sim$ 8 hours. 
A similar dependence on monitoring duration has been noticed for RQQs, 
where the chances of INOV detection increases to $\sim 24\%$ if the 
monitoring is done for $\sim$6-7 hours (de Diego et al. 1998;
Carini et al. 2007). It is 
conceivable that the INOV duty cycle of $\sim 9 - 14\%$ estimated here
for RIQs may go up marginally when longer monitoring sessions become
possible, but this is unlikely to alter our main conclusions.

The simplest explanation for the difference in variability is to assume 
that all quasars do possess nuclear radio jets, with the differences 
in both observed INOV DCs and amplitudes explained by different Doppler 
boosting arising from both different viewing angles and different 
shock velocities (e.g., Gopal-Krishna et al. 2003). 
Another possible explanation for the low INOV DC would involve dilution 
of jet's optical emission by the expected accretion 
disc emission; however, this is unlikely to be important here, 
since all the RIQs in our 
sample lie at small redshifts ($z < 1.5 $; Table 1) whereas the bulk of the 
 continuum emission of the disk would be expected to arise in the far-UV
and would only move 
in the R-band for $z > 2$ (e.g., Bachev et al. 2005; Carini et al. 2007).
Furthermore, since all but one of these RIQs have $M_B < -24.5$, the 
chance of the host
galaxies' stellar continua diluting the INOV when compared to the other
classes of QSOs is also very small.

Were the INOV properties of AGNs to depend primarily on the beamed
radio emission, one would expect stronger and more frequent INOV 
for our sample of RIQs, as compared to that found for RQQs. Such an
expectation would be in tune with the widely held notion that RIQs are
Doppler boosted counterparts of RQQs, as inferred from their high 
brightness temperature ($T_{B}$), prominent radio cores and low 
extended fluxes at radio
wavelengths, measured using milli-arcsecond resolution radio images
available for a few RIQs 
(Section 1). The INOV data reported here do not accord well these broad
expectations, since we have found the INOV of RIQs to be as mild as 
that known for RQQs.

Thus, the RIQ PG0007+106 (III Zw 2) remains  the only well 
established case for which a blazar-like, large and rapid flux 
variability has been detected (both at radio and optical wavelengths, 
Sect.\ 1). In particular, the source was found by Jang $\&$ Miller 
(1997) to vary by up to 0.1 mag on each the two nights they monitored
it for 4-hour duration. Interestingly, the observed high level of activity 
would not have been anticipated, given that this object was found to 
show a very low optical polarization: $P_{op} = 0.28 \pm 0.19\%$, 
(Berriman et al. 1990).
The RIQ III Zw 2 underwent a slow fading by $\sim$0.8 mag between 1978 
to 1981, after which it displayed rapid flares of increasing amplitude, 
brightening by 0.92 magnitude in 27 days during November 1985 and then 
fading by 0.95 mag in the following week (Pica et al. 1988). Such large 
variability, along with the observed INOV (see above) is clearly 
characteristic of OVV type blazars. In the longer term, its optical light 
curve has shown a four-fold variation in 25 years (Salvi et al. 2002) 
and, likewise, a 20-fold brightening within 4 years has been recorded at 
radio wavelengths (Aller et al. 1985). All these variability properties 
firmly place this object in the blazar category (Kukula et al. 1998;  
Aller et al. 1999; Ter\"asranta et al. 2005). In contrast, none of the 
10 RIQs in our representative sample qualify for the blazar classification, 
on the basis of the currently available data summarized above. 

The present observations are also useful for measuring any long-term optical 
variability (LTOV) occurring on month-like or longer timescales. For the 
vast majority of RIQs, which are represented by the 
present sample and which are hosted by luminous (very likely elliptical)
galaxies, we find the optical variability to be very common on 
month/year-like time scales, with typical amplitude approaching 0.1-mag 
level in R-band. This RIQ specific result is in accord with the findings
reported by Webb \& Malkan (2000) for more common AGN types. For roughly half 
the AGNs they found optical variability amplitudes of 0.1 -- 0.2 mag (rms)
on month-like time scale. A similar pattern is reported by Barvainis et al.
(2005), based on their 10-epoch VLA radio variability survey of a large 
sample consisting of RQQs, RIQs and RLQs.
They found no statistical difference between the degrees of radio variability 
displayed by these three major subclasses of quasars. Thus, both intranight 
and long-term optical variability properties, together with the other 
afore-cited evidence from the literature, demonstrate that as a class
RIQs display much lower activity levels than do blazars, which would be 
rather surprising if they are a strongly relativistically beamed subset 
of RQQs, a possibility widely discussed in the literature.

The possibility that these RIQ sources possess intrinsically moderate radio
jets and are not strongly Doppler boosted (in either radio or optical band) 
therefore must be considered.  In that case either roughly comparable
amounts of jet fluctuations, or even disc dominated variability, could
explain the similar INOV and LTOV properties of RIQs and RQQs. The low
optical polarizations seen for RIQs in our sample (Table 1) also seem
consistent with a lack of strong relativistic beaming, though the case 
of III Zw 2 discussed above calls for caution in reaching this inference. 
A key observational difference between these RIQs and the RQQs is the 
predominance of flat or inverted radio spectra for the former group, which 
is generally not the case for RQQs (e.g., Kukula et al. 1998) 
and which was argued to imply (at least somewhat) boosted jets in the RIQs. 

One conceivable explanation then for the relative lack of INOV in RIQs would be 
that, because of bending of their jets on the innermost scales, 
their optically emitting inner portions are misdirected and hence concealed 
from us (despite the beaming) and only the more extended radio emitting outer 
parts of the jets happen to be pointed towards us. Verifying this alternative 
would require sensitive VLBI observations. 
However, an implication of this scenario is that one would expect to find some
cases of blazar-like optical variability in radio-quiet AGN whose radio
emitting (but not optical emitting) sections of the jet are misaligned from
us. The few searches made so far for radio-quiet BL Lacs have either been
negative (e.g., Stocke et al. 1990; Londish et al. 2007) or have produced 
a only a small fraction of candidate BL Lacs that have low, but not extremely
low, upper limits to their radio fluxes (Collinge et al. 2005).
Finally, it may be noted that despite the vast 
increase in the intranight monitoring data, as reported here, the total 
number of nights used for their monitoring is still modest (42 nights) when compared 
with those devoted to blazars and even RQQs. Since a few convincing cases of RIQs showing 
modest intranight or internight variability have nonetheless been found, 
it would be worthwhile to continue such observations in the search for 
examples of blazar-like strong INOV activity among RIQs, similar to that 
detected so far only for the nearest known RIQ, III Zw 2.

\section*{Acknowledgments}
AG would like to thank Dr. Vijay Mohan for help during 
observations at IGO and for a preliminary discussion on data analysis.
The authors wish to acknowledge the support rendered by the staff of IAO and 
CREST at IIA and the IUCAA-Girawali Observatory (IGO), IUCAA.
This research has made use of NASA/IPAC Extragalactic Database (NED), 
which is operated by the Jet Propulsion Laboratory, California Institute 
of Technology, under contract with National Aeronautics and Space 
Administration.


\newpage
\begin{table*}
\caption{Sample of 10 RIQs used in the present study}
\begin{tabular}{ccccccccccc}\\
\hline
IAU Name&RA(J2000)&Dec.(J2000)&{\it B}& $M_{B}$& z &$P_{op}^{\P}(\%)$&$P_{5GHz}^\S$&$\alpha_{radio}$&$log R^{*\dag}$&References\\
 (1)    &  (2)    & (3)       &  (4)  &  (5)   &(6)&  (7)       & (8)     &  (9)            & (10)          &  (11)    \\
\hline

J0748+2200$^*$ & 07 48 15.4 &$+$22 00 59.6&17.18 &$-$27.2 & 1.059&  -    &$2.9\times10^{25}$$^c$&  -             & 1.106 & (1)\\
J0832+3707$^*$ & 08 32 25.3 &$+$37 07 36.7&16.61 &$-$22.1 & 0.091&  -    &$1.1\times10^{23}$$^c$&$-$0.50         & 1.142 & (1)\\
J0836+4426$^*$ & 08 36 58.9 &$+$44 26 02.4&15.09 &$-$25.9 & 0.249&  -    &$1.7\times10^{24}$$^c$&$+$0.14         & 0.422 & (1)\\
J0907+5515$^*$ & 09 07 43.6 &$+$55 15 12.5&17.81 &$-$25.2 & 0.645&  -    &$2.4\times10^{25}$$^c$&  -             & 1.788 & (1)\\
J1259+3423$^*$ & 12 59 48.7 &$+$34 23 22.8&17.05 &$-$28.0 & 1.375&  0.65$^a$ &$6.0\times10^{25}$$^c$& $+$0.06        & 1.090 & (2)\\
J1312+3515     & 13 12 17.7 &$+$35 15 20.8&15.64 &$-$24.6 & 0.184&  0.31$^b$ &$2.7\times10^{24}$$^d$&   0.00$^\ddag$ & 1.312 & (3)\\
J1336+1725$^*$ & 13 36 01.9 &$+$17 25 14.0&16.23 &$-$26.5 & 0.554&  0.18$^b$ &$2.7\times10^{25}$$^e$&$-$0.31$^\ddag$ & 1.330 & (3)\\
J1539+4735$^*$ & 15 39 34.8 &$+$47 35 31.0&15.81 &$-$27.7 & 0.772&  0.90$^b$ &$2.6\times10^{25}$$^e$&$+$0.19$^\ddag$ & 1.439 & (3)\\
J1701+5149     & 17 01 25.0 &$+$51 49 20.0&15.49 &$-$25.8 & 0.292&  0.54$^b$ &$1.1\times10^{23}$$^f$&$-$0.2$^*    $ & 0.914 & (3)\\
J1719+4804$^*$ & 17 19 38.3 &$+$48 04 13.0&14.60 &$-$29.8 & 1.083&  0.40$^b$ &$2.6\times10^{26}$$^g$&$+$0.49$^\ddag$ & 0.863 & (3)\\
\hline
\end{tabular}
{$^\P$} References for the optical polarization: (a) Stockmann et al. (1984) 
(b) Berriman et al. (1990)\\ 
{$\S$} The radio luminosity at 5 GHz (W/Hz/Sr) is calculated using high resolution fluxes available from the
literature. The references are: (c) Becker, White \& Helfand (1995); 
(d) Falcke, Patnaik \& Sherwood (1996a); (e) Kellermann et al. (1989); \
(f) Blundell \& Beasly (1998); (g) Helmboldt et al (2007).
If measurements were at a different frequency,
then the luminosities were converted into 5 GHz luminosities
assuming the core spectral index $\alpha_{C} =0.0$ ($S_{\nu} \propto \nu^{\alpha}$) 
wherever the spectral index of the source was not known.\\
{{$\dag${\it $R^{*}$}} is the {\it K}-corrected ratio of the 5 GHz to 2500
\AA~flux densities (Stocke et al.\ 1992)$;$ references for the radio
fluxes are V\'{e}ron-Cetty \& V\'{e}ron (2006), NVSS (Condon et al.\ 1998)
and FIRST (Becker, White \& Helfand 1995).  \\
{$\ddag$} From Falcke, Sherwood \& Patnaik (1996b) which were 
derived using the quasi-simultaneous observations at 2.7 and 10.0 GHz  
with $S_{\nu} \propto \nu^{\alpha}$ while the remainder have been calculated 
using non-simultaneous obsevartions available in NED.\\
{$^*$} From Kukula et al.\ (1998)  using observations of the core at 8.4 and 4.8 GHz.\\}
Reference in Col. (11): (1) Wang et al. (2006); (2) Carini et al. (2007);
(3) Falcke, Sherwood \& Patnaik (1996b) 

\end{table*}

\begin{table*}
\caption{Summary of observations and the INOV paramters }
\begin{tabular}{ccccccccccc}\\
\hline
Source   &    DATE  &Tel. used&Filter &Duration&$N_{points}$&$\sigma $&$\psi$ &   $ C_{eff}$  & Status$^*$& References  \\
         &  dd.mm.yy&         &       & (hours)&           & (\%)    &  (\%)  &               &           &             \\
\hline
J0748+2200  & 23.01.07 &  ST        & R   &  6.4  & 23   &   0.17  &       2.3&          2.58 & PV$\S$ & (a) \\
J0748+2200  & 19.02.07 &  ST        & R   &  5.9  & 22   &   0.26  &       0.7&          0.40 & N   &   (a) \\
J0748+2200  & 29.01.08 & IGO        & R   &  4.9  & 17   &   0.09  &       0.5&          0.76 & N   &   (a) \\
J0748+2200  & 30.01.08 & IGO        & R   &  5.5  & 18   &   0.10  &       1.4&          5.10 & V$\S$& (a) \\
            &          &            &     &       &      &         &          &               &     &       \\
J0832+3707  & 23.01.07 & HCT        & R   &  4.6  & 27   &   0.15  &       1.2&          1.76 & N   &   (a) \\
J0832+3707  & 21.02.07 &  ST        & R   &  4.2  & 19   &   0.15  &       1.5&          1.58 & N   &   (a) \\
J0832+3707  & 10.03.07 & IGO        & R   &  4.0  & 08   &   0.16  &       0.8&          0.63 & N   &   (a) \\
J0832+3707  & 11.03.07 & IGO        & R   &  4.0  & 08   &   0.30  &       0.6&          0.28 & N   &   (a) \\
            &          &            &     &       &      &         &          &               &     &       \\
J0836+4426  & 22.01.07 &  ST        & R   &  5.2  & 22   &   0.12  &       0.9&          0.74 & N   &   (a) \\
J0836+4426  & 10.02.07 & IGO        & R   &  3.9  & 13   &   0.35  &       1.1&          1.08 & N   &   (a) \\
J0836+4426  & 09.03.07 & IGO        & R   &  4.2  & 14   &   0.26  &       2.1&          1.54 & N   &   (a) \\
            &          &            &     &       &      &         &          &               &     &       \\
J0909+5515  & 04.02.08 & IGO        & R   &  4.2  & 22   &   0.20  &       1.2&          0.68 & N   &   (a) \\
J0909+5515  & 05.02.08 & IGO        & R   &  6.4  & 11   &   0.17  &       0.5&          0.36 & N   &   (a) \\
            &          &            &     &       &      &         &          &               &     &       \\
J1259+3423  & 11.05.98 &            &V/R  &  5.8  &      &         &          &               & N   &   (b) \\
J1259+3423  & 12.05.98 &            &V/R  &  6.9  &      &         &          &               & N   &   (b) \\
J1259+3423  & 19.04.07 &  ST        & R   &  5.0  & 19   &   0.22  &       1.0&          0.81 & N   &   (a) \\
J1259+3423  & 20.04.07 &  ST        & R   &  5.9  & 25   &   0.25  &       1.9&          2.50 & PV  &   (a) \\
J1259+3423  & 24.04.07 &  ST        & R   &  3.9  & 20   &   0.25  &       1.1&          0.59 & N   &   (a) \\
            &          &            &     &       &      &         &          &               &     &       \\
J1312+3515  & 27.02.00 &            &V/R  &  4.0  &      &         &          &               & N   &   (b) \\
J1312+3515  & 29.02.00 &            &V/R  &  6.0  &      &         &          &               & N   &   (b) \\
J1312+3515  & 02.03.00 &            &V/R  &  5.0  &      &         &          &               & N   &   (b) \\
J1312+3515  & 08.03.99 &  ST        & R   &  6.7  & 39   &         &          &               & N   &   (c) \\
J1312+3515  & 01.04.01 &  ST        & R   &  4.6  & 32   &         &          &               & N   &   (c) \\
J1312+3515  & 02.04.01 &  ST        & R   &  5.2  & 41   &         &          &               & N   &   (c) \\
            &          &            &     &       &      &         &          &               &     &       \\
J1336+1725  & 11.04.05 &  ST        & R   &  7.0  & 27   &   0.18  &       0.9&          0.65 & N   &   (a) \\
J1336+1725  & 08.05.05 &  ST        & R   &  3.9  & 16   &   0.31  &       1.4&          1.95 & N   &   (a) \\
J1336+1725  & 13.04.08 &  ST        & R   &  6.9  & 17   &   0.19  &       1.3&          1.02 & N   &   (a) \\
            &          &            &     &       &      &         &          &               &     &       \\
J1539+4735  & 21.04.01 &            & R   &  5.0  &      &   1.20  &          &               & N   &   (d) \\
J1539+4735  & 27.05.09 &  ST        & R   &  5.8  & 26   &   0.31  &       1.6&          0.99 & N   &   (a) \\
J1539+4735  & 02.06.09 &  ST        & R   &  6.5  & 28   &   0.22  &       1.4&          0.48 & N   &   (a) \\
J1539+4735  & 14.06.09 &  ST        & R   &  4.4  & 20   &   0.26  &       1.2&          0.53 & N   &   (a) \\
            &          &            &     &       &      &         &          &               &     &       \\
J1701+5149  & 04.06.99 &            &V/R  &  5.8  &      &         &          &               & N   &   (b) \\
J1701+5149  & 05.06.99 &            &V/R  &  4.9  &      &         &          &               & N   &   (b) \\
J1701+5149  & 06.06.99 &            &V/R  &  5.1  &      &         &          &               & N   &   (b) \\
J1701+5149  & 07.06.99 &            &V/R  &  5.7  &      &         &          &               & N   &   (b) \\
J1701+5149  & 26.06.02 &            & R   &  5.3  &      &   1.00  &          &          3.1  & V   &   (d) \\
            &          &            &     &       &      &         &          &               &     &       \\
J1719+4804  & 01.08.94 &            & V/R &  3.3  &      &   0.90  &          &          2.6  & V   &   (e) \\
J1719+4804  & 05.08.94 &            & V/R &  3.8  &      &   1.10  &          &               & N   &   (e) \\
J1719+4804  & 11.06.98 &            & R   &  7.6  &      &   0.90  &          &               & N   &   (d) \\
J1719+4804  & 29.04.06 &  ST        & R   &  4.5  & 22   &   0.15  &       0.8&          1.62 & N   &   (a) \\
J1719+4804  & 30.04.06 &  ST        & R   &  5.0  & 26   &   0.13  &       0.6&          1.20 & N   &   (a) \\
J1719+4804  & 30.05.06 &  ST        & R   &  4.6  & 24   &   0.17  &       0.6&          0.66 & N   &   (a) \\
\hline
\end{tabular}

{$^*$ V = Variable; N = Non-variable; PV = Probable Variable}\\
$\S$ see text for explanation (section 4.2)\\
{References for INOV data: (a) Present work; (b) Carini et al.\ (2007); (c) Sagar et al.\ (2004); (d) Jang (2005); (e)
 Jang \& Miller (1995)} 

\end{table*}

\clearpage
\newpage
\begin{table}
\caption{Positions and magnitudes of the RIQs and the comparison stars used
in the present study$^*$.}
\begin{tabular}{lccccc}\\
\hline
Source & RA(J2000) &Dec.(J2000) &  {\it B} & {\it R}  & {\it B-R}  \\
      &           &            & (mag)    & (mag)   \\\hline
J0748+2200 & 07{$^h$}48{$^m$}15.{$^s$}43 & $+$22{$^\circ$}00{$^\prime$}59{$^{\prime\prime}$}.6 & 16.25 & 15.59 & 0.66  \\
S1         & 07{$^h$}48{$^m$}03.{$^s$}57 & $+$22{$^\circ$}03{$^\prime$}48{$^{\prime\prime}$}.8 & 16.56 & 16.47 & 0.09  \\
S2         & 07{$^h$}48{$^m$}01.{$^s$}31 & $+$22{$^\circ$}00{$^\prime$}10{$^{\prime\prime}$}.7 & 15.64 & 15.17 & 0.47  \\
S3         & 07{$^h$}47{$^m$}58.{$^s$}65 & $+$22{$^\circ$}01{$^\prime$}34{$^{\prime\prime}$}.9 & 16.31 & 15.83 & 0.48  \\
J0832+3707 & 08{$^h$}32{$^m$}25.{$^s$}35 & $+$37{$^\circ$}07{$^\prime$}36{$^{\prime\prime}$}.7 & 15.42 & 15.57 & -0.15 \\
S1         & 08{$^h$}32{$^m$}39.{$^s$}63 & $+$37{$^\circ$}12{$^\prime$}17{$^{\prime\prime}$}.7 & 16.85 & 15.16 & 1.69  \\
S2         & 08{$^h$}32{$^m$}39.{$^s$}78 & $+$37{$^\circ$}11{$^\prime$}56{$^{\prime\prime}$}.5 & 17.02 & 15.59 & 1.43  \\
S3         & 08{$^h$}32{$^m$}29.{$^s$}94 & $+$37{$^\circ$}11{$^\prime$}30{$^{\prime\prime}$}.2 & 16.05 & 15.34 & 0.71  \\
J0836+4426 & 08{$^h$}36{$^m$}58.{$^s$}91 & $+$44{$^\circ$}26{$^\prime$}02{$^{\prime\prime}$}.4 & 15.46 & 15.44 & 0.02  \\
S1         & 08{$^h$}37{$^m$}07.{$^s$}62 & $+$44{$^\circ$}25{$^\prime$}57{$^{\prime\prime}$}.9 & 15.87 & 15.05 & 0.82  \\
S2         & 08{$^h$}37{$^m$}14.{$^s$}23 & $+$44{$^\circ$}21{$^\prime$}33{$^{\prime\prime}$}.5 & 17.27 & 15.73 & 1.54  \\
S3         & 08{$^h$}36{$^m$}54.{$^s$}69 & $+$44{$^\circ$}22{$^\prime$}50{$^{\prime\prime}$}.1 & 17.22 & 16.11 & 1.11  \\
S4         & 08{$^h$}37{$^m$}10.{$^s$}75 & $+$44{$^\circ$}22{$^\prime$}14{$^{\prime\prime}$}.7 & 15.60 & 14.29 & 1.31  \\
J0907+5515 & 09{$^h$}07{$^m$}43.{$^s$}64 & $+$55{$^\circ$}15{$^\prime$}12{$^{\prime\prime}$}.5 & 17.27 & 17.38 & -0.11 \\
S1         & 09{$^h$}07{$^m$}33.{$^s$}94 & $+$55{$^\circ$}18{$^\prime$}10{$^{\prime\prime}$}.7 & 17.56 & 16.69 & 0.87  \\
S2         & 09{$^h$}07{$^m$}33.{$^s$}98 & $+$55{$^\circ$}13{$^\prime$}30{$^{\prime\prime}$}.2 & 17.89 & 16.47 & 1.42  \\
S3         & 09{$^h$}07{$^m$}42.{$^s$}29 & $+$55{$^\circ$}11{$^\prime$}33{$^{\prime\prime}$}.0 & 17.66 & 16.64 & 1.02  \\
J1259+3423 & 12{$^h$}59{$^m$}48.{$^s$}79 & $+$34{$^\circ$}23{$^\prime$}22{$^{\prime\prime}$}.8 & 17.31 & 16.06 & 1.25  \\
S1         & 12{$^h$}59{$^m$}43.{$^s$}91 & $+$34{$^\circ$}23{$^\prime$}22{$^{\prime\prime}$}.3 & 17.35 & 15.25 & 2.1   \\
S2         & 13{$^h$}00{$^m$}18.{$^s$}24 & $+$34{$^\circ$}21{$^\prime$}46{$^{\prime\prime}$}.9 & 18.54 & 15.71 & 2.83  \\
S3         & 13{$^h$}00{$^m$}20.{$^s$}69 & $+$34{$^\circ$}23{$^\prime$}06{$^{\prime\prime}$}.5 & 16.39 & 14.55 & 1.84  \\
J1336+1725 & 13{$^h$}36{$^m$}02.{$^s$}00 & $+$17{$^\circ$}25{$^\prime$}13{$^{\prime\prime}$}.1 & 17.11 & 15.84 & 1.27  \\
S1         & 13{$^h$}35{$^m$}59.{$^s$}44 & $+$17{$^\circ$}31{$^\prime$}20{$^{\prime\prime}$}.6 & 17.26 & 15.31 & 1.95  \\
S2         & 13{$^h$}36{$^m$}21.{$^s$}27 & $+$17{$^\circ$}26{$^\prime$}36{$^{\prime\prime}$}.3 & 16.84 & 14.89 & 1.95  \\
S3         & 13{$^h$}36{$^m$}17.{$^s$}66 & $+$17{$^\circ$}28{$^\prime$}05{$^{\prime\prime}$}.7 & 17.68 & 14.71 & 2.97  \\
S4         & 13{$^h$}35{$^m$}31.{$^s$}90 & $+$17{$^\circ$}21{$^\prime$}31{$^{\prime\prime}$}.6 & 16.00 & 14.42 & 1.58  \\
J1539+4735 & 15{$^h$}39{$^m$}34.{$^s$}80 & $+$47{$^\circ$}35{$^\prime$}31{$^{\prime\prime}$}.4 & 16.25 & 15.16 & 1.09  \\
S1         & 15{$^h$}39{$^m$}11.{$^s$}40 & $+$47{$^\circ$}30{$^\prime$}56{$^{\prime\prime}$}.9 & 16.21 & 14.64 & 1.57  \\
S2         & 15{$^h$}39{$^m$}13.{$^s$}01 & $+$47{$^\circ$}30{$^\prime$}18{$^{\prime\prime}$}.7 & 16.89 & 15.28 & 1.61  \\
S3         & 15{$^h$}39{$^m$}42.{$^s$}35 & $+$47{$^\circ$}35{$^\prime$}07{$^{\prime\prime}$}.1 & 16.74 & 15.42 & 1.32  \\
J1719+4804 & 17{$^h$}19{$^m$}38.{$^s$}25 & $+$48{$^\circ$}04{$^\prime$}12{$^{\prime\prime}$}.5 & 14.98 & 14.25 & 0.73  \\
S1         & 17{$^h$}19{$^m$}14.{$^s$}96 & $+$48{$^\circ$}03{$^\prime$}56{$^{\prime\prime}$}.5 & 17.50 & 16.84 & 0.66  \\
S2         & 17{$^h$}19{$^m$}18.{$^s$}22 & $+$48{$^\circ$}08{$^\prime$}02{$^{\prime\prime}$}.6 & 15.40 & 14.62 & 0.78  \\
S3         & 17{$^h$}19{$^m$}13.{$^s$}68 & $+$48{$^\circ$}04{$^\prime$}50{$^{\prime\prime}$}.5 & 17.50 & 16.67 & 0.83  \\
S4         & 17{$^h$}18{$^m$}56.{$^s$}21 & $+$48{$^\circ$}06{$^\prime$}44{$^{\prime\prime}$}.9 & 15.23 & 14.44 & 0.79  \\
\hline
\hline
\end{tabular}

$^*$ From the United States Naval Observatory-B catalogue. Note that these magnitudes are only
accurate up to 0.3 mag and the uncertainity in position is 2{$^{\prime\prime}$}(Monet et al.\ 2003). 
\end{table}
\clearpage

\newpage
\begin{figure*}
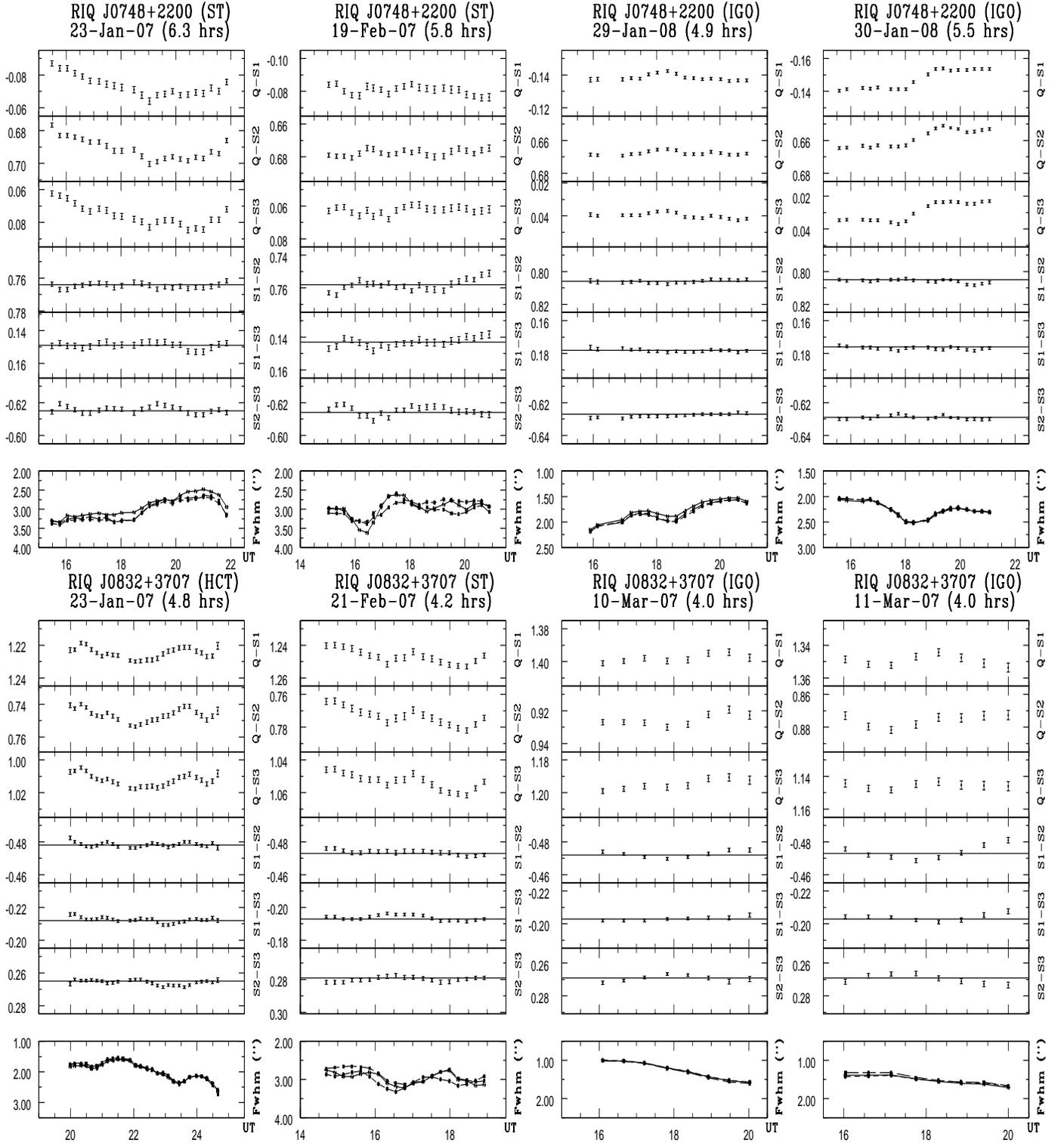

\hspace*{-1.0cm}
\hbox{
\includegraphics[height=10.0cm,width=04.5cm]{fig_J0748+2200_ST_23jan07.epsi}
\includegraphics[height=10.0cm,width=04.5cm]{fig_J0748+2200_ST_19feb07.epsi}
\includegraphics[height=10.0cm,width=04.5cm]{fig_J0748+2200_IGO_29jan08.epsi}
\includegraphics[height=10.0cm,width=04.5cm]{fig_J0748+2200_IGO_30jan08.epsi}
}
\vspace*{1cm}
\hspace*{-1.0cm}
\hbox{
\includegraphics[height=10.0cm,width=04.5cm]{fig_J0832+3707_HCT_23jan07.epsi}
\includegraphics[height=10.0cm,width=04.5cm]{fig_J0832+3707_ST_21feb07.epsi}
\includegraphics[height=10.0cm,width=04.5cm]{fig_J0832+3707_IGO_10mar07.epsi}
\includegraphics[height=10.0cm,width=04.5cm]{fig_J0832+3707_IGO_11mar07.epsi}
}
\caption{The intranight optical DLCs of the RIQs monitored in the present study. 
For each night, the source, the telescope used,  
the date, and the duration of monitoring are given.
The upper 3 panels show the DLCs of the source relative to 3 
comparison stars while the attached lower 3 panels shows the star-star DLCs, where 
the solid horizontal lines mark the mean for each steady star-star DLC. 
The bottom panel gives the plots of seeing variation for the night, 
based on 3 stars monitored along the RIQ on the same CCD frame. }
\label{fig:1}
\end{figure*}
\clearpage
\begin{figure*}
\hspace*{-1.0cm}
\hbox{
\includegraphics[height=10.0cm,width=04.5cm]{fig_J0836+4426_ST_22jan07.epsi}
\includegraphics[height=10.0cm,width=04.5cm]{fig_J0836+4426_IGO_10feb07.epsi}
\includegraphics[height=10.0cm,width=04.5cm]{fig_J0836+4426_IGO_09mar07.epsi}
\includegraphics[height=10.0cm,width=04.5cm]{fig_J0907+5515_IGO_04feb08.epsi}
}
\vspace*{1cm}
\hspace*{-1.0cm}
\hbox{
\includegraphics[height=10.0cm,width=04.5cm]{fig_J0907+5515_IGO_05feb08.epsi}
\includegraphics[height=10.0cm,width=04.5cm]{fig_J1259+3423_ST_19apr07.epsi}
\includegraphics[height=10.0cm,width=04.5cm]{fig_J1259+3423_ST_20apr07.epsi}
\includegraphics[height=10.0cm,width=04.5cm]{fig_J1259+3423_ST_24apr07.epsi}
}
\begin{center}
{{\bf Figure~\ref{fig:1}}. \textit {continued}}
\end{center}
\end{figure*}
\clearpage
\newpage

\begin{figure*}
\hspace*{-1.0cm}
\hbox{
\includegraphics[height=10.0cm,width=04.5cm]{fig_J1336+1725_ST_11apr05.epsi}
\includegraphics[height=10.0cm,width=04.5cm]{fig_J1336+1725_ST_08may05.epsi}
\includegraphics[height=10.0cm,width=04.5cm]{fig_J1336+1725_ST_13apr08.epsi}
\includegraphics[height=10.0cm,width=04.5cm]{fig_J1539+4735_ST_27may09.epsi}
}
\vspace*{1cm}
\hspace*{-1.0cm}
\hbox{
\includegraphics[height=10.0cm,width=04.5cm]{fig_J1539+4735_ST_02jun09.epsi}
\includegraphics[height=10.0cm,width=04.5cm]{fig_J1539+4735_ST_14jun09.epsi}
\includegraphics[height=10.0cm,width=04.5cm]{fig_J1719+4804_ST_29apr06.epsi}
\includegraphics[height=10.0cm,width=04.5cm]{fig_J1719+4804_ST_30apr06.epsi}
}
\begin{center}
{{\bf Figure~\ref{fig:1}}. \textit {continued}}
\end{center}
\end{figure*}
\clearpage

\newpage
\begin{figure*}
\hspace*{-1.0cm}
\hbox{
\includegraphics[height=10.0cm,width=04.5cm]{fig_J1719+4804_ST_30may06.epsi}

}
\begin{center}
{{\bf Figure~\ref{fig:1}}. \textit {continued}}
\end{center}
\end{figure*}

\newpage
\begin{figure*}
\includegraphics[height=22.0cm,width=16.0cm]{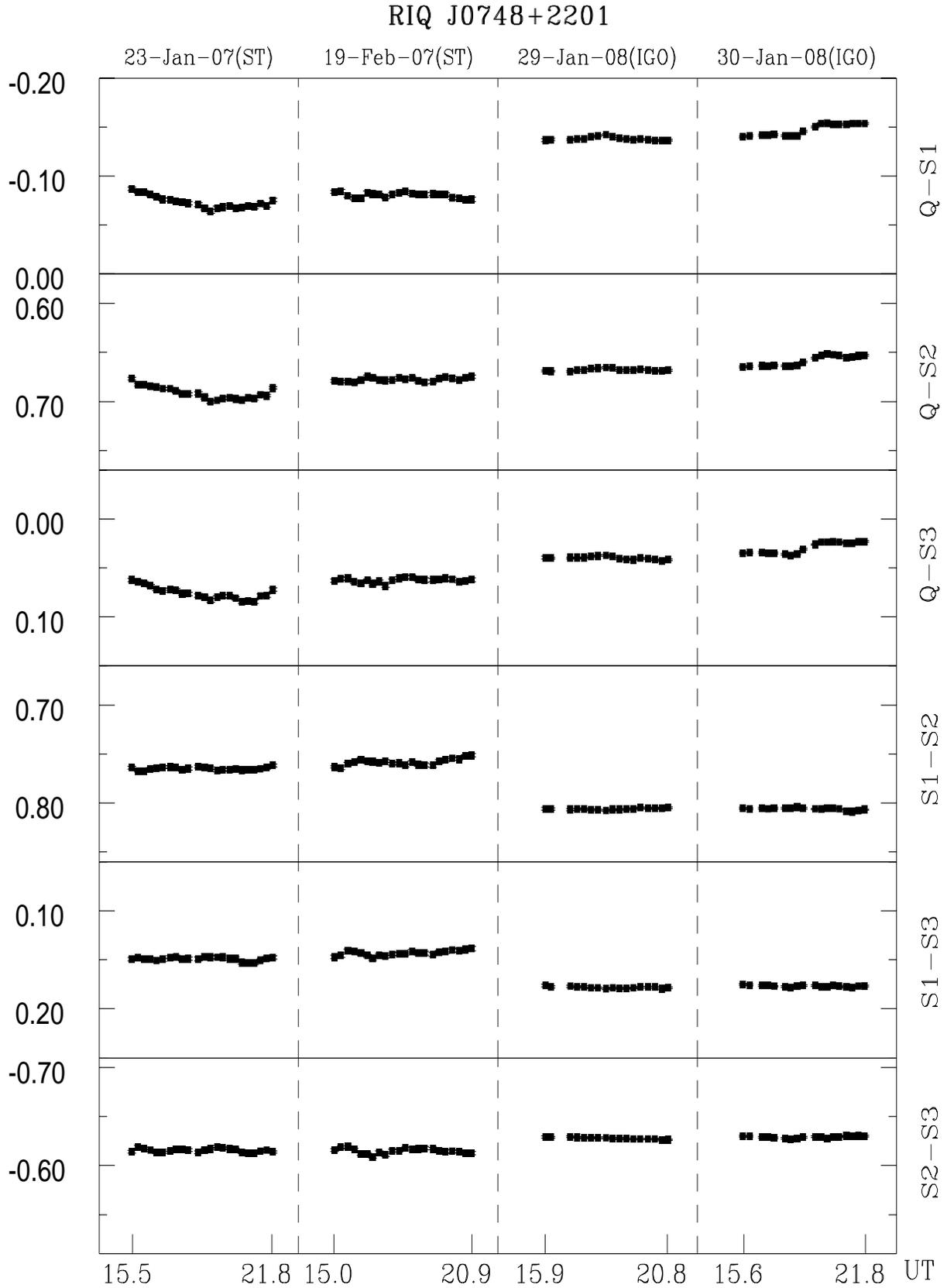}
\caption{The LTOV DLCs of RIQs monitored
in the present sample} 
\label{fig:2}
\end{figure*}
\clearpage

\newpage
\begin{figure*}
\includegraphics[height=22.0cm,width=16.0cm]{ltov_J0832+3707.epsi}
\begin{center}
{{\bf Figure~\ref{fig:2}}. \textit {continued}}
\end{center}
\end{figure*}
\clearpage

\newpage
\begin{figure*}
\includegraphics[height=22.0cm,width=12.0cm]{ltov_J0836+4426.epsi}
\begin{center}
{{\bf Figure~\ref{fig:2}}. \textit {continued}}
\end{center}
\end{figure*}
\clearpage

\newpage
\begin{figure*}
\includegraphics[height=22.0cm,width=08.0cm]{ltov_J0907+5515.epsi}
\begin{center}
{{\bf Figure~\ref{fig:2}}. \textit {continued}}
\end{center}
\end{figure*}

\newpage
\begin{figure*}
\includegraphics[height=22.0cm,width=12.0cm]{ltov_J1259+3423.epsi}
\begin{center}
{{\bf Figure~\ref{fig:2}}. \textit {continued}}
\end{center}
\end{figure*}

\newpage
\begin{figure*}
\includegraphics[height=22.0cm,width=12.0cm]{ltov_J1336+1725.epsi}
\begin{center}
{{\bf Figure~\ref{fig:2}}. \textit {continued}}
\end{center}
\end{figure*}
\clearpage
\newpage
\begin{figure*}
\includegraphics[height=22.0cm,width=12.0cm]{ltov_J1539+4735.epsi}
\begin{center}
{{\bf Figure~\ref{fig:2}}. \textit {continued}}
\end{center}
\end{figure*}

\newpage
\begin{figure*}
\includegraphics[height=22.0cm,width=12.0cm]{ltov_J1719+4804.epsi}
\begin{center}
{{\bf Figure~\ref{fig:2}}. \textit {continued}}
\end{center}
\end{figure*}
\clearpage
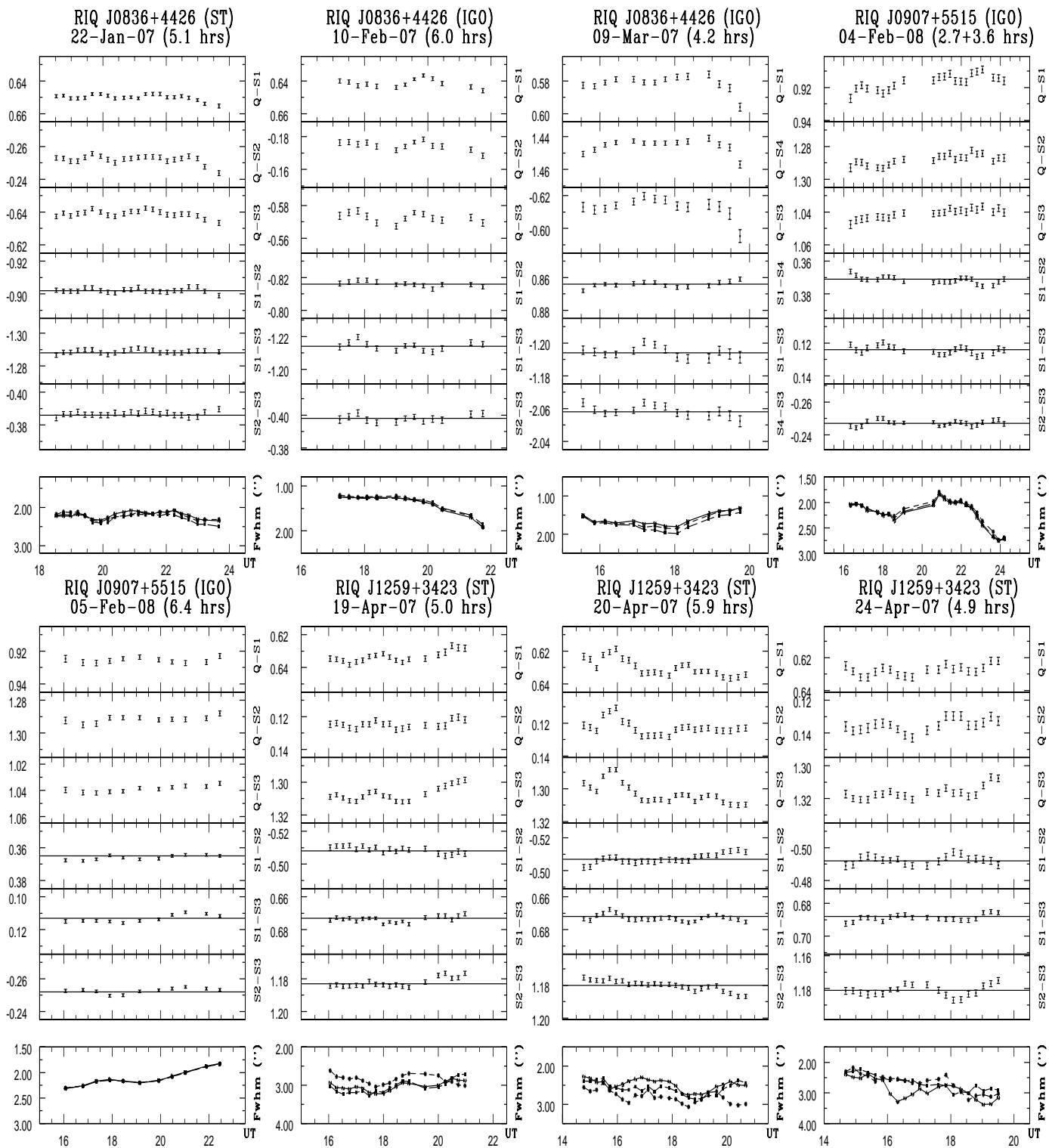
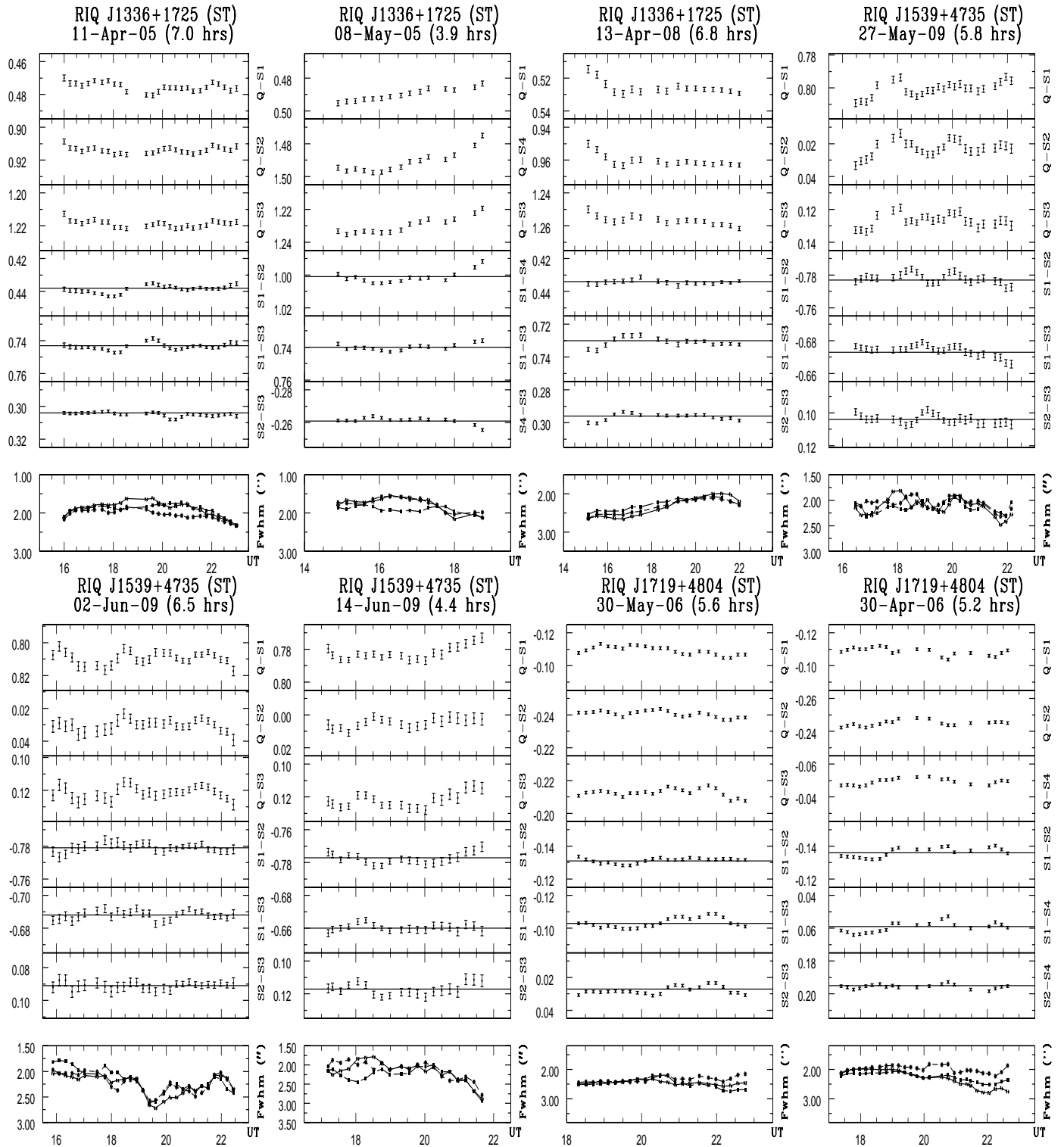
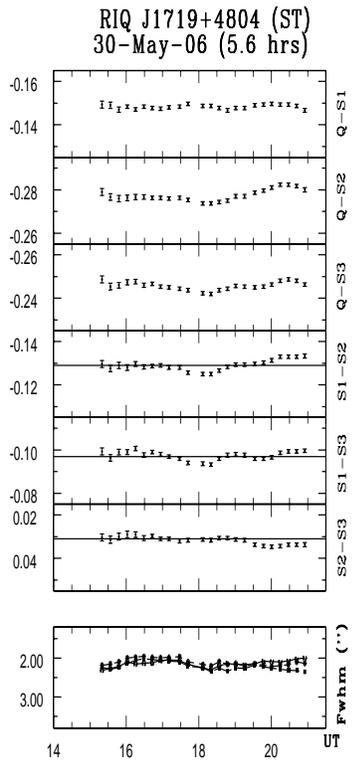
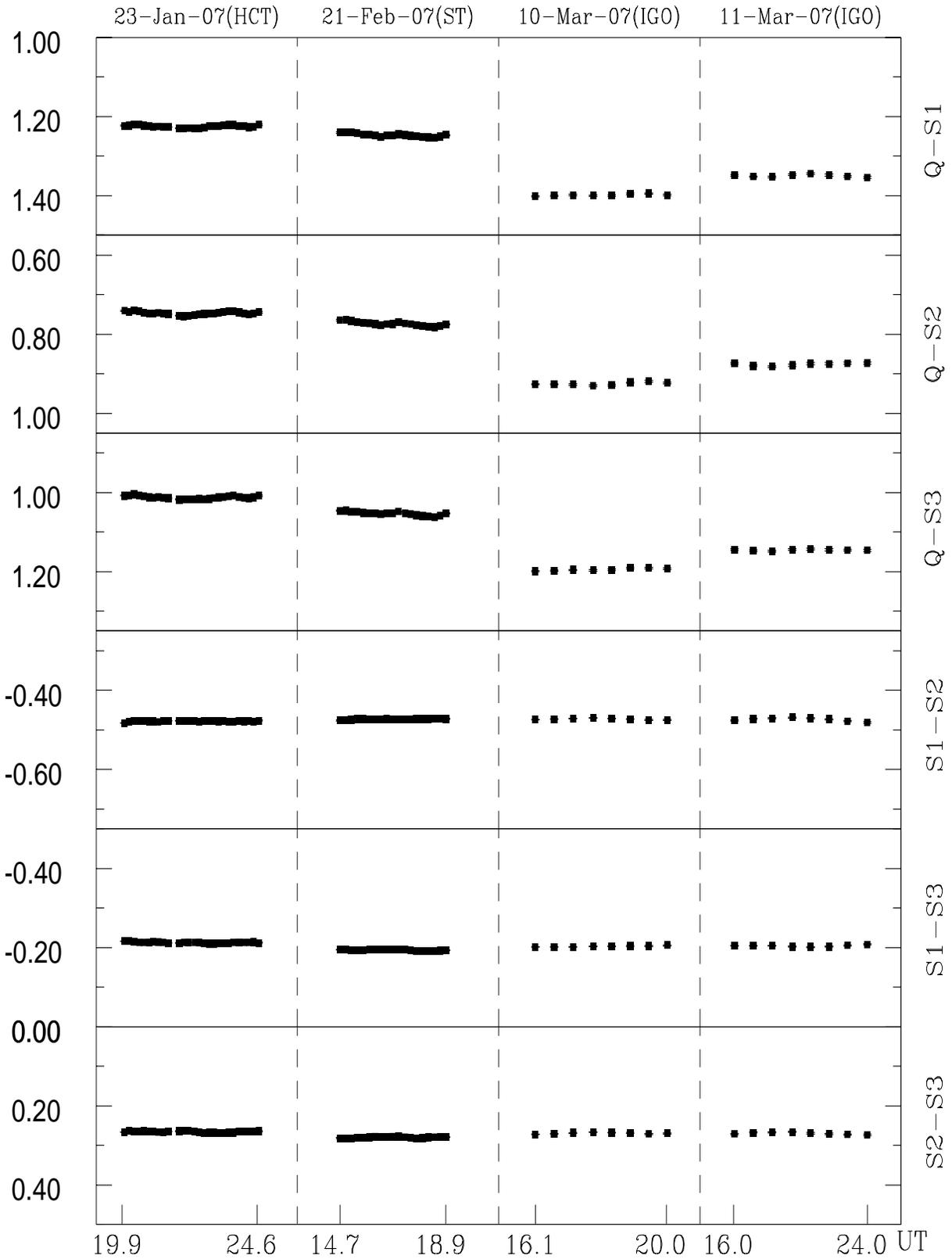
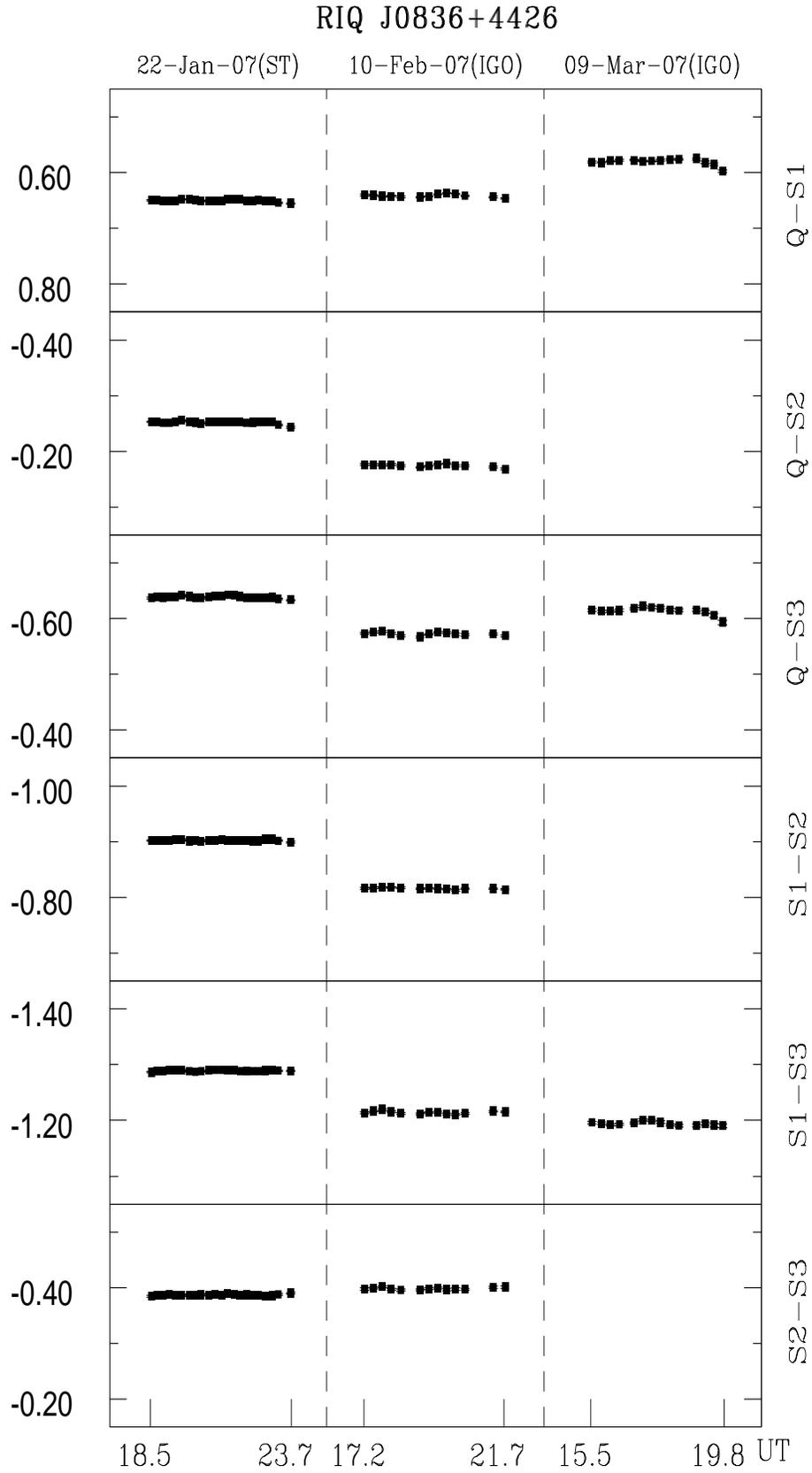
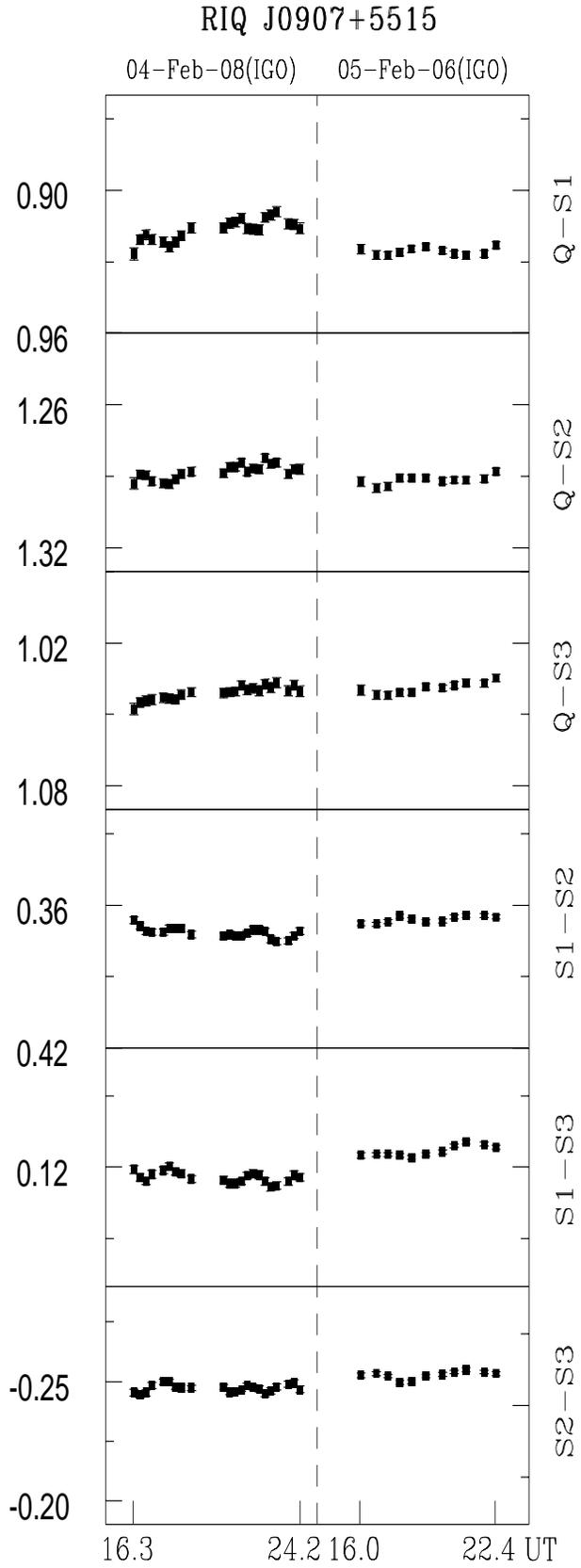
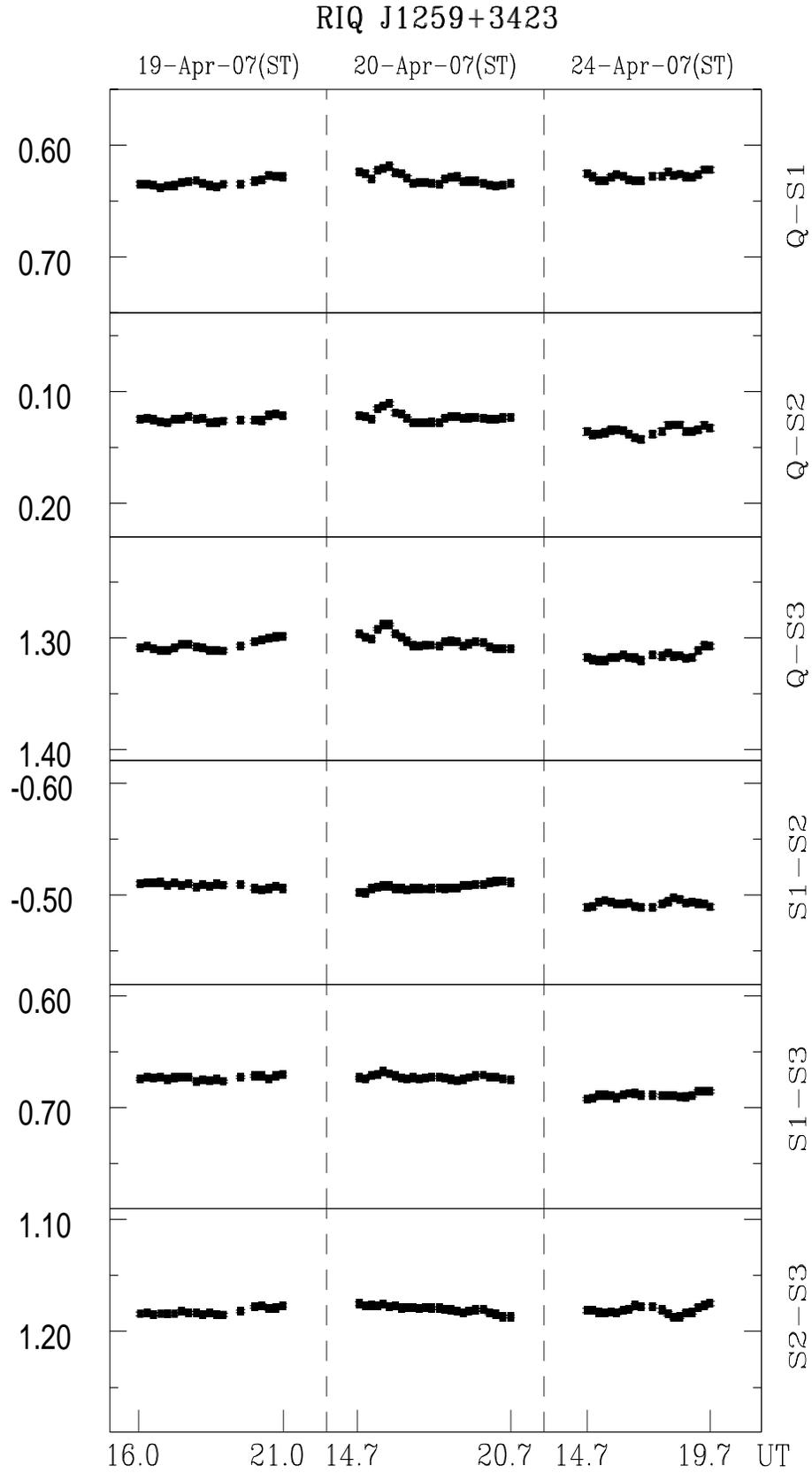
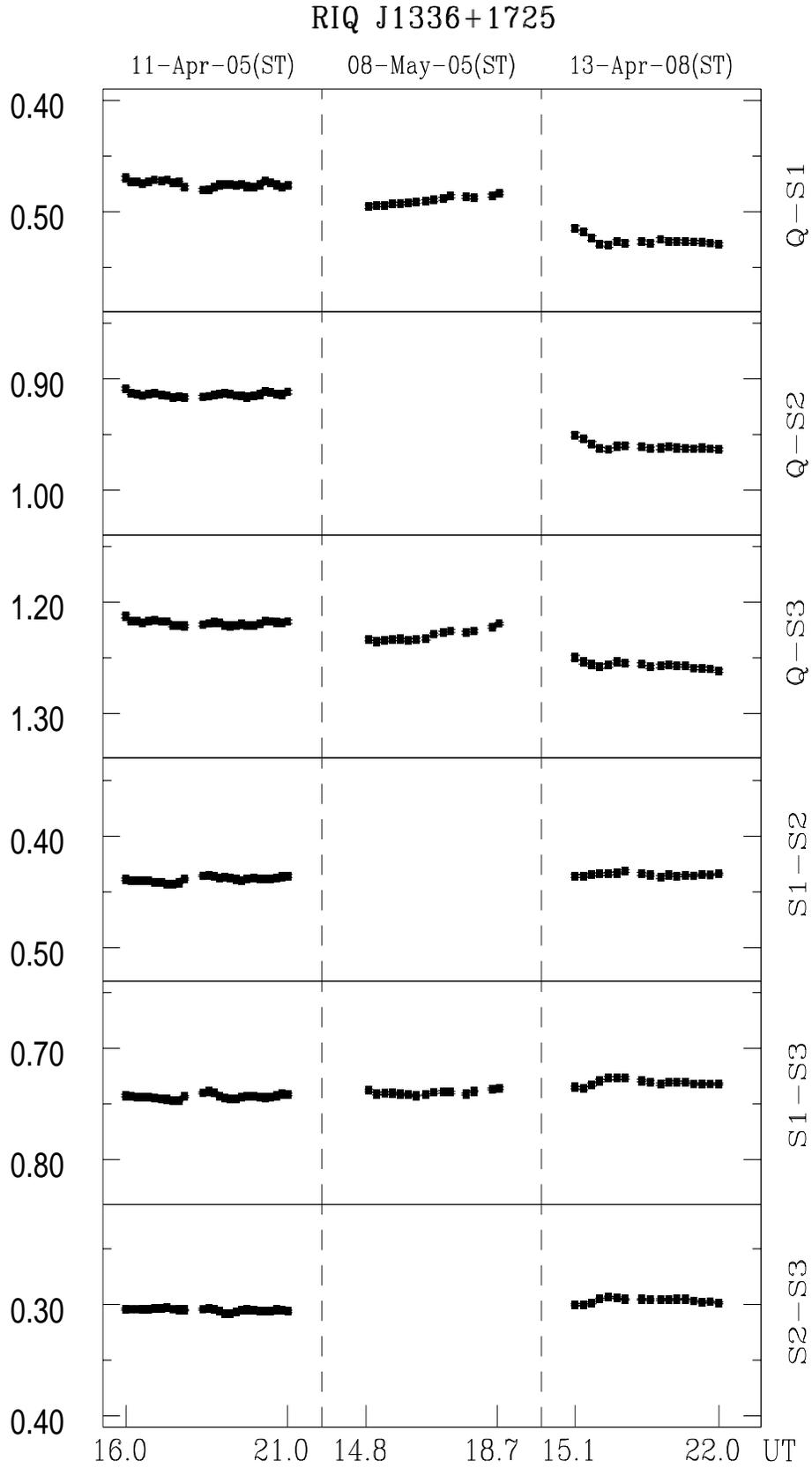
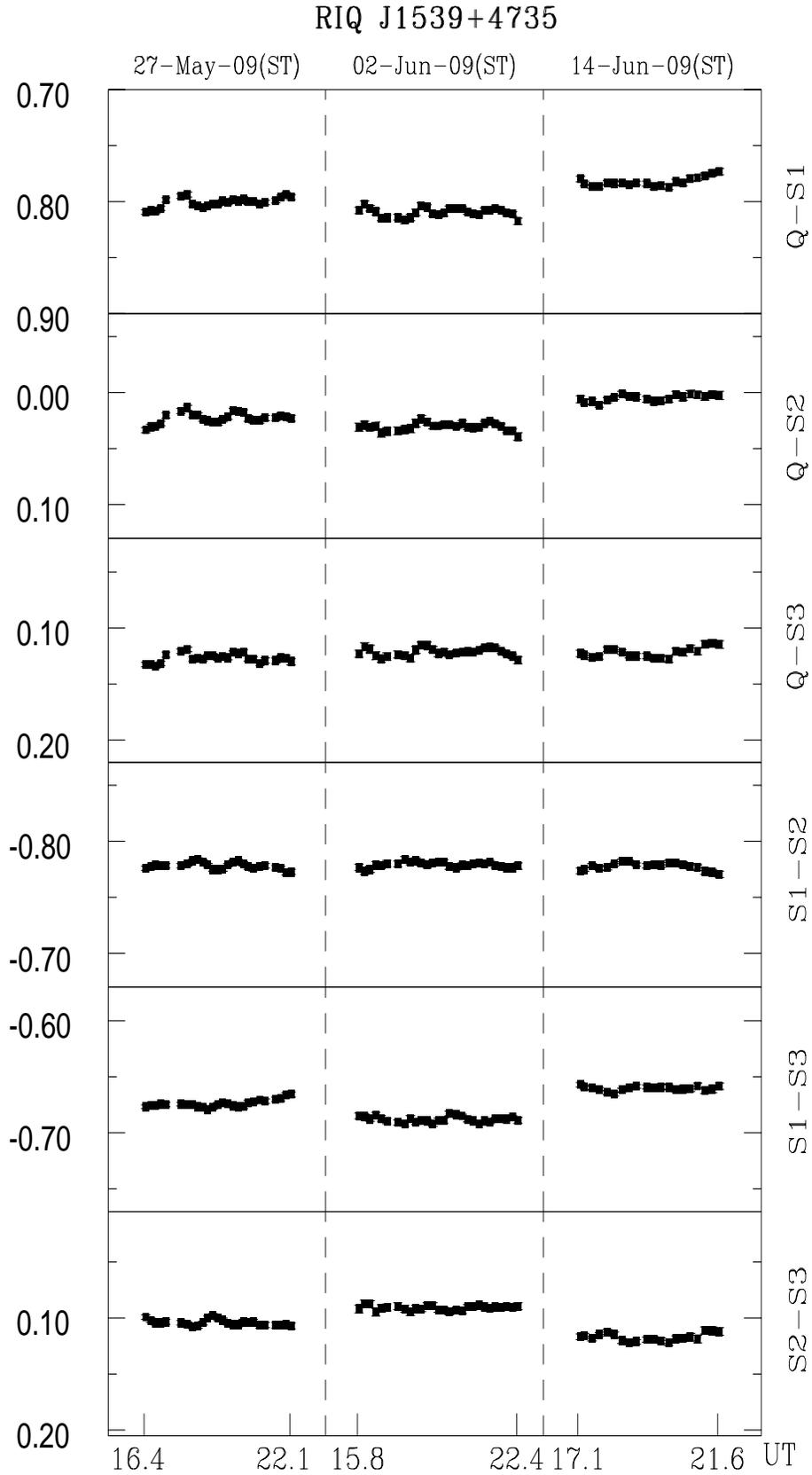
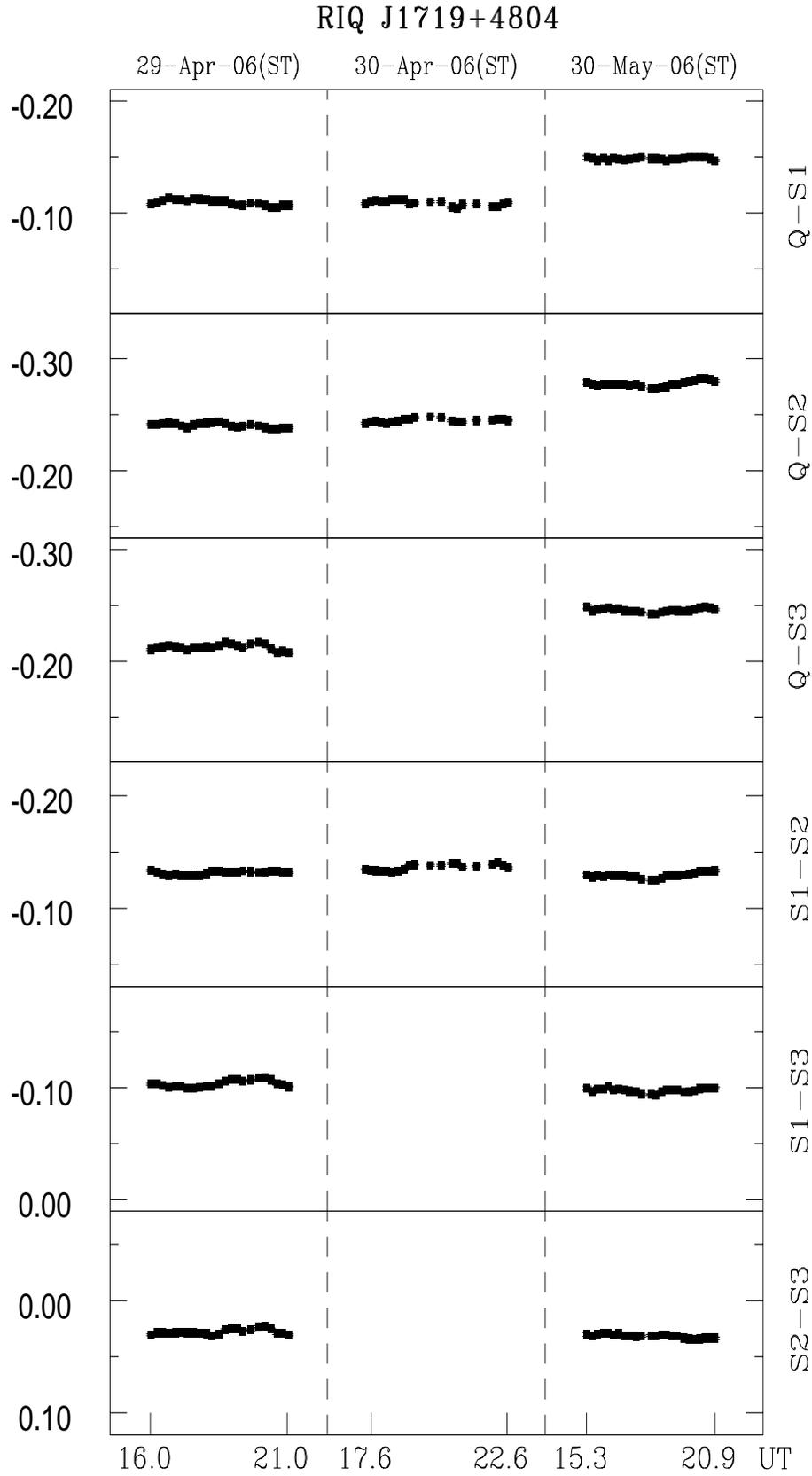
\end{document}